\documentclass[aps,prx,reprint,superscriptaddress,floatfix,twocolumn,nofootinbib]{revtex4-2}
\usepackage{graphicx}
\usepackage{amsmath}
\usepackage{amsfonts}
\usepackage{bm}
\usepackage{color}
\usepackage{indentfirst}
\usepackage[normalem]{ulem}
\begin{document}
\title{Dynamics and Rheology of Ring-Linear Blend Semidilute Solutions in Extensional Flow: Single Molecule Experiments}

\author{Yuecheng Zhou}
\altaffiliation{Current address: Department of Chemistry, Stanford University, Stanford, California 94305, USA}
\affiliation{Department of Materials Science and Engineering, University of Illinois at Urbana-Champaign, Urbana, Illinois 61801, USA}
\affiliation{Beckman Institute for Advanced Science and Technology, University of Illinois at Urbana-Champaign, Urbana, Illinois 61801, USA}

\author{Charles D. Young}
\affiliation{Beckman Institute for Advanced Science and Technology, University of Illinois at Urbana-Champaign, Urbana, Illinois 61801, USA}
\affiliation{Department of Chemical and Biomolecular Engineering, University of Illinois at Urbana-Champaign, Urbana, Illinois 61801, USA}

\author{Kathryn E. Regan}
\affiliation{Department of Physics, University of San Diego, San Diego, California 92110, USA}

\author{Megan Lee}
\affiliation{Department of Physics, University of San Diego, San Diego, California 92110, USA}

\author{Sourya Banik}
\affiliation{Department of Chemical Engineering, Texas Tech University, Lubbock, Texas 79409, USA}

\author{Dejie Kong}
\affiliation{Department of Chemical Engineering, Texas Tech University, Lubbock, Texas 79409, USA}

\author{Gregory B. McKenna}
\affiliation{Department of Chemical Engineering, Texas Tech University, Lubbock, Texas 79409, USA}

\author{Rae M. Robertson-Anderson}
\affiliation{Department of Physics, University of San Diego, San Diego, California 92110, USA}

\author{Charles E. Sing}
\affiliation{Beckman Institute for Advanced Science and Technology, University of Illinois at Urbana-Champaign, Urbana, Illinois 61801, USA}
\affiliation{Department of Chemical and Biomolecular Engineering, University of Illinois at Urbana-Champaign, Urbana, Illinois 61801, USA}

\author{Charles M. Schroeder}
\email{To whom correspondence should be addressed: cms@illinois.edu}
\affiliation{Department of Materials Science and Engineering, University of Illinois at Urbana-Champaign, Urbana, Illinois 61801, USA}
\affiliation{Beckman Institute for Advanced Science and Technology, University of Illinois at Urbana-Champaign, Urbana, Illinois 61801, USA}
\affiliation{Department of Chemical and Biomolecular Engineering, University of Illinois at Urbana-Champaign, Urbana, Illinois 61801, USA}

\date{\today}
\begin{abstract}
Ring polymers exhibit unique flow properties due to their closed chain topology. Despite recent progress, we have not yet achieved a full understanding of the nonequilibrium flow behavior of rings in nondilute solutions where intermolecular interactions greatly influence chain dynamics. In this work, we directly observe the dynamics of DNA rings in semidilute ring-linear polymer blends using single molecule techniques. We systematically investigate ring polymer relaxation dynamics from high extension and transient and steady-state stretching dynamics in planar extensional flow for a series of ring-linear blends with varying ring fraction. Our results show multiple molecular sub-populations for ring relaxation in ring-linear blends, as well as large conformational fluctuations for rings in steady extensional flow, even long after the initial transient stretching process has subsided. We further quantify the magnitude and characteristic timescales of ring conformational fluctuations as a function of blend composition. Interestingly, we find that the magnitude of ring conformational fluctuations follows a non-monotonic response with increasing ring fraction, first increasing at low ring fraction and then substantially decreasing at large ring fraction in ring-linear blends. A unique set of ring polymer conformations are observed during the transient stretching process, which highlights the prevalence of molecular individualism and supports the notion of complex intermolecular interactions in ring-linear polymer blends. In particular, our results suggest that transient intermolecular structures form in ring-linear blends due to a combination of direct forces due to linear chains threading through open rings and indirect forces due to hydrodynamic interactions; these combined effects lead to large conformational fluctuations of rings over distributed timescales. Taken together, our results provide a new molecular understanding of ring polymer dynamics in ring-linear blends in nonequilibrium flow. 
\end{abstract}

\maketitle
\section{Introduction}
	Ring polymers have a topologically closed structure with no beginning or end. Due to their unique properties,  ring polymers have captured the attention of rheologists and soft materials scientists for decades \cite{McLeish2002}. Beyond their intriguing macromolecular structures, ring polymers are of practical importance in several disciplines. In nature, mitochondrial DNA and plasmid DNA generally occur in cyclic forms \cite{Taanman1999}. Genome organization in cell nuclei has been modeled as a melt of nonconcatenated ring polymers, which represents the simplest model where reptation is suppressed due to topological constraints \cite{Halverson2014}. Prior work has examined the molecular threading of linear chains through macrocyclic oligomeric rings \cite{Deutman2008}, and recent advances in synthetic organic chemistry have enabled the synthesis of cyclic rings using olefin metathesis \cite{Edwards2019}. In addition, synthetic ring polymers have been used to generate transient materials with triggered degradation properties \cite{Feinberg2018}, thereby providing promising new routes towards the development of fully recyclable synthetic materials \cite{Lloyd2019,Puskas2015,Puskas2012}.

	The flow properties of ring polymer solutions and melts have long been a topic of interest in the community. Early efforts to understand the flow behavior of ring polymer melts focused on synthetic polystyrene and polybutadiene rings using shear rheology \cite{Roovers1985, McKenna1987, Roovers1988, Mckenna1989}. In general, ring polymer melts exhibit a smaller zero-shear viscosity, $\eta_0$, and a larger recoverable compliance, $J_e^0$, in the terminal flow regime compared to linear melt counterparts \cite{Mckenna1989}. Ring polymer melts also show a surprising dependence of zero-shear viscosity on molecular weight, $M_w$ \cite{Mckenna1989}. In particular, $\eta_0$ for ring polymer melts is approximately one-half the value of linear polymer melt counterparts below the critical entanglement molecular weight, $M_e$. For higher molecular weight rings, $\eta_0$ generally shows a smaller power-law scaling exponent compared to the commonly observed $\eta_0 \propto M_w^{3.4}$ scaling for linear polymers \cite{McKenna1987, Halverson2011b, Pasquino2013a}, albeit for ring melts with nominal entanglement densities relative to linear polymers of fewer than 15 entanglements per chain. Moreover, ring polymer melts exhibit no rubbery plateau and show a faster terminal relaxation that significantly contrasts with linear polymer melts undergoing stress relaxation \cite{Kapnistos2008a, Doi2017, Obukhov1994, Ge2016}. However, the rheological response of ring polymers is highly susceptible to linear chain contamination \cite{McKenna1986a,Roovers1988}. For example, it was reported that even a small amount of linear chains (as small as 0.07$\%$ by volume) drastically increases the zero-shear viscosity of ring polymer melts and causes the rubbery plateau to reappear \cite{Kapnistos2008a}. However, it has been challenging to precisely quantify trace amounts of linear polymers in ring polymer melts, with subsequent experiments showing some differences in rheological response with quantitatively different linear chain content \cite{Doi2017}.
	
	A major challenge in experimental characterization of ring polymers lies in preparing high purity ring samples that are essentially free of linear chains. To this end, advances in chromatography techniques (liquid chromatography at the critical condition, LCCC) have led to improved separation of rings from linear polymers \cite{Lee2000}, thereby enabling experimental studies of the linear viscoelasticity \cite{Kapnistos2008a,Yan2016}, non-linear shear rheology \cite{Yan2016}, and extensional flow properties of LCCC-purified ring melts \cite{Huang2019a} and ring-linear blends \cite{Borger2020}. Nevertheless, prior work has shown that ring polymer samples obtained by post-polymerization cyclization of linear chains followed by chromatographic purification invariably contain small amounts of linear chains that affect rheological measurements despite rigorous purification using the LCCC method \cite{Doi2015,mckenna2021}. Such observations strongly motivate the need to understand the flow behavior of ring-linear polymer blends \cite{Borger2020}.
	
	The equilibrium properties of ring-linear blends have been studied for concentrated polymer solutions and melts \cite{Chapman2012a, Iyer2007, Halverson2012}. Molecular architecture and blend composition both play major roles in determining the conformation and size of ring polymers in these systems. In ring-linear blend melts, increasing the fraction of linear chains leads to an increase in the radius of gyration, $R_g$, for rings and a drastic decrease in the ring polymer diffusion coefficient in the blend \cite{Halverson2012, Iyer2007}. Interestingly, the excluded volume exponent, $\nu$, for ring polymers remains relatively constant in good solvent conditions for ring polymer solutions in the presence of linear polymers up to 5-10$\%$ linear chains in solution \cite{Gartner2019}. 
	
	Experimental and computational studies have further shown that molecular topology significantly alters the diffusion coefficient of ring polymers. Single molecule experiments on ring DNA show that rings diffuse $\approx$1.3$\times$ faster than linear polymers in concentrated ring DNA solutions \cite{Robertson2007c}. Remarkably, rings diffuse $\approx$10$\times$ slower than linear polymers when the background matrix contains concentrated linear DNA molecules \cite{Robertson2007a, Subramanian2008}. A marked difference in the molecular weight scaling dependence of polymer diffusion was observed for rings diffusing in concentrated solutions of rings versus linear polymers, implying different underlying mechanisms of chain diffusion in ring versus linear backgrounds \cite{Robertson2007a}. Ring polymers were also reported to exhibit heterogeneous multimodal diffusion when the concentration of the background linear polymer solution increases well into the entangled regime \cite{Habuchi2010}. In ring-linear blends, the center-of-mass diffusion behavior of ring polymers shows rich dynamics due to mixed chain topologies. For example, the diffusion coefficient of rings was found to decrease monotonically upon increasing the fraction of linear polymers from 0 to 50$\%$ in a concentrated ring-linear polymer blend \cite{Chapman2012a}. This phenomenon emerges when the background blend concentration approaches the critical entanglement concentration for linear chains, $c_e$, and becomes enhanced when the blend is entangled at concentrations above $c_e$. 
	
	In order to elucidate the key physical features of ring-linear topological constraints, several theoretical models have been proposed to explain the slow-down of ring polymer dynamics in entangled solutions of ring-linear polymer blends. Constraint release (CR) or restricted reptation (RR) of background linear polymers, where rings relax through amoebae-like motion in fixed obstacles formed by background linear chains, was used to model the stress relaxation of rings in melts of ring-linear blends \cite{Graessley1982, Klein1986}. The once-threaded model (R1) was subsequently developed \cite{Mills1987}, wherein rings diffuse along a threaded linear chain. Here, `threading' refers to one or more linear chains in the background matrix penetrating into an open ring conformation, resulting in a significant decrease in ring polymer diffusion \cite{Yang2010}. Despite these conceptual advances, the actual diffusion mechanism is not fully understood and may be composed of elements from several of these models \cite{Tsalikis2016}. 
	
	The majority of prior work on ring-linear polymer blends has focused on the equilibrium properties such as ring polymer size or ring diffusion in concentrated solutions or melts \cite{Robertson2007a, Robertson2007c, Subramanian2008, Habuchi2010, Halverson2012, Chapman2012a, Ge2016, Kruteva2017}. However, the flow behavior of ring-linear blends is of paramount importance for processing applications. In 2019, nonlinear extensional rheology was performed on LCCC-purified ring polystyrene melts, with results showing unexpected increases in extensional viscosity at low stretch rates \cite{Huang2019a} due to topological linking of rings \cite{o2020topological}. In 2020, nonlinear extensional rheology combined with molecular dynamics (MD) simulations and ex situ small angle neutron scattering was used to probe the threading-unthreading behavior of ring and linear polymers in melts of ring-linear blends \cite{Borger2020}.
	
	Whereas shear and extensional rheology and small angle neutron scattering (SANS) characterization provides useful insight into ring-linear polymer blends, such bulk-level experiments only provide information on  ensemble-averaged properties, which tends to obscure dynamics at the molecular level. Recent advances in single molecule fluorescence microscopy (SMFM) and automated flow control enable the direct observation of polymer dynamics under nonequilibrium flow conditions \cite{Schroeder2018}. SMFM can be used to identify and characterize molecular sub-populations of polymer chains that adopt different transient conformations or show molecular individualism in flow \cite{Perkins1997, Smith1999, Soh2018}. In recent years, SMFM has been used to study the dynamics of linear polymers in dilute solution large amplitude oscillatory extensional flow (LAOE) \cite{Zhou2016,Zhou2016b} and in semidilute unentangled solutions in extensional flow \cite{Hsiao2017,Samsal2017,Young2019a}. Single polymer dynamics was also used to study the relaxation of linear polymers in entangled solutions \cite{Zhou2018}, revealing unexpectedly heterogeneous dynamics. 
	
	The flow properties of ring polymers in dilute solutions were recently studied using single molecule experiments and simulations \cite{Li2015, Hsiao2016a, Weiss2017, Young2019}. In dilute solution extensional flow, ring polymers show reduced molecular individualism during transient stretching and a shifted coil-stretch transition (CST) compared to linear chains due to combined effects of a closed, constrained molecular topology and intramolecular hydrodynamic interactions between the two ring strands \cite{Li2015, Hsiao2016a}. The dynamics of single ring polymers in the flow-gradient plane of shear was recently studied using a combination of SMFM with a custom shear flow apparatus and Brownian dynamics simulations \cite{Tu2020}, where it was observed that the probability of chain extension in the flow direction was qualitatively different for rings compared to linear chains in shear flow \cite{Tu2020}.
	
	Single molecule studies of rings have also been extended to semidilute polymer solutions. Recently, \citet{Zhou2019} studied the extensional flow dynamics of ring polymers in semidilute solutions of linear polymers near the overlap concentration $c^*$, which is defined as the concentration at which linear polymer molecules begin to overlap and interpenetrate at equilibrium and therefore defines the transition between the dilute and semidilute unentangled regimes \cite{Graessley1982}. In steady extensional flow, rings exhibit large conformational fluctuations in semidilute solutions which was attributed to the transient threading of linear polymers through open rings stretching in flow \cite{Zhou2019}. Remarkably, such large conformational fluctuations of rings emerged at extremely low concentrations of background linear polymers (0.025 $c^*$). Overall, these studies showcase the ability of single molecule techniques to reveal the effects of molecular architecture on the flow dynamics of ring polymers. 
	
\begin{figure}[t]
 			\includegraphics[scale=1]{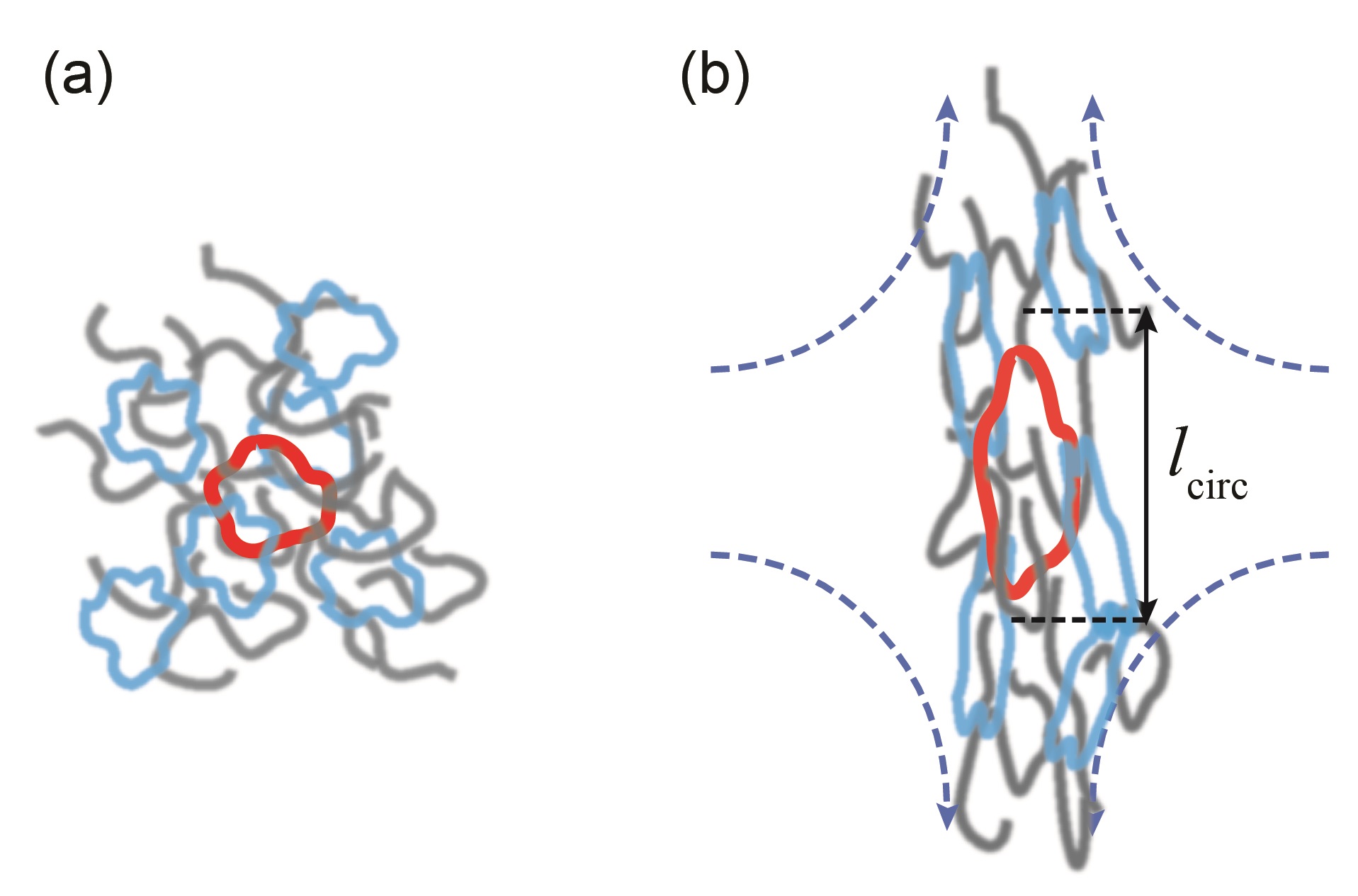}
  			\caption{Schematic of ring-linear DNA polymer blends. Fluorescently labeled tracer ring DNA molecules (45 kbp, red) are uniformly dissolved in a semidilute blend solution of rings (blue) and linear chains (gray). Ring polymer dynamics are investigated at (a) equilibrium or zero-flow conditions and in (b) planar extensional flow. The transient molecular stretch of ring polymers $l_\mathrm{{circ}}$ is directly measured using SMFM.}
			\label{Schematic}
		\end{figure}

	In this work, we study the relaxation and transient stretching dynamics of ring DNA molecules in semidilute ring-linear polymer blends using SMFM (Fig. \ref{Schematic}). Fluorescently labeled ring DNA molecules (45 kbp) are introduced into ring-linear DNA blend solutions of equivalent molecular weight. In this way, we study the flow dynamics of three different ring-linear blends containing 17$\%$ rings (R) and 83$\%$ linear (L) polymers by mass (17$\%$ R-83$\%$ L), 50$\%$ R-50$\%$ L, and 83$\%$ R-17$\%$ L. Results are compared to ring dynamics in semidilute background solutions of purely linear polymers (0$\%$ R-100$\%$ L). Experimental results are complemented by Brownian dynamics (BD) simulations of ring-linear blend solutions. Essential points of comparison are included in this manuscript, and a detailed simulation study for ring dynamics is presented in a companion paper \cite{young2020dynamics}. Our results show that the magnitude of ring conformational fluctuations exhibits a non-monotonic response with increasing ring fraction in blends, first increasing at low ring fraction and then substantially decreasing for large ring fractions in the blend ($>80\%$ R). Conformational fluctuations are quantified in terms of an average fluctuation magnitude $\langle \delta \rangle$ and a characteristic fluctuation timescale using autocorrelation analysis. Interestingly, we identify a unique set of molecular conformations during the transient stretching process for ring polymers, suggesting complex intermolecular interactions between rings and polymer chains in the ring-linear blend solution. We further determine average fractional extension in semidilute ring-linear blends; in all cases, rings show an overall decreased fractional extension for ring-linear blends in extensional flow compared to dilute solution rings or pure linear chains in semidilute solutions. Taken together, experimental and computational results show that transient ring conformations in flow are driven by ring-linear threading interactions and long-range intermolecular hydrodynamic interactions (HI) in semidilute solutions.

\section{Materials and methods}
\textbf{A. Preparation of 45 kbp ring and linear DNA}

	Double-stranded 45 kbp ring and linear DNA molecules are prepared via replication of fosmids in \textit{Escherichia coli}, followed by extraction and purification, as previously described \cite{Laib2006a, Robertson2006a, Robertson2007c}. Briefly, circular DNA molecules are extracted from bacterial cell cultures using alkaline lysis, followed by renaturation of the cloned DNA by treatment with an acidic detergent solution. Genomic DNA and cellular debris precipitate are removed by centrifugation, and supercoiled DNA molecules are converted to relaxed circular conformations via treatment with Topoisomerase-I (New England Biolabs) \cite{Robertson2006a,Peddireddy2020}. To generate linear DNA, restriction endonucleases are used to specifically cut the double stranded DNA ring backbone at precisely one location. Ring and linear polymer samples are treated with RNase A to remove contaminating RNA, and excess protein is removed by phenol-chloroform extraction followed by dialysis. Finally, DNA samples are concentrated by a second isopropanol precipitation, and the molecular topology and concentration are determined using gel electrophoresis \cite{Laib2006a, Robertson2007c}. In general, the concentration of the prepared (stock) ring and linear DNA solutions is $\approx$500 $\mu$g/mL. The purity of ring DNA samples is further characterized using single molecule visualization. Here, small amounts of DNA samples are taken from each batch and fluorescently labeled, as described below. Fluorescently labeled samples are then diluted to a concentration of approximately 5$\times 10^{-4}$ ng/$\mu$L in the imaging buffer and introduced into a cross-slot microfluidic device for imaging. In this way, single DNA molecules are stretched in extensional flow, which allows for direct observation and classification of ring or linear topology. This process is repeated for an ensemble of at least 200 molecules for reliable statistics, enabling quantification of the ring-linear fraction in each sample. \\
	
\textbf{B. Preparation of semidilute ring-linear blends}

	 For all experiments, the total polymer concentration in ring-linear DNA blends was maintained at 50 $\mu$g/mL, which corresponds to $\approx$1 $c^*$ for 45 kbp linear DNA \cite{Zhou2018}. In order to prepare ring-linear DNA blends with varying ring fractions, we first calculate the mass of ring and linear DNA required at different target ring and linear DNA compositions with a desired total volume of 5 mL, which is a typical sample volume used for single molecule imaging and viscosity measurements. Next, based on the stock 45 kbp ring DNA concentration and its ring-linear content, a working volume of 45 kbp ring DNA solution is prepared and heated to 65 $^\circ$C for 10 minutes, followed by snap cooling to 0 $^\circ$C. A working volume of 45 kbp linear DNA is also prepared following a similar procedure. Both working volumes are then concentrated to $\approx$50 $\mu$L using a MiVac Quattro concentrator (Genevac, UK). Next, the concentrated working volumes of 45 kbp ring DNA and linear DNA are mixed and diluted with viscous buffer solution containing 30 mM Tris/Tris-HCl (pH 8.0), 2 mM EDTA, 5 mM NaCl and sucrose (66.3 $\%$ w/w) to a final sample volume of 5 mL. Prepared semidilute ring-linear blends then undergo a gentle rotational mixing procedure for approximately 4 hours at $22.5$ $^\circ$C to ensure sample homogeneity, followed by overnight gentle rotational mixing at 4 $^\circ$C. Ring-linear DNA blends with varying ring fraction and their corresponding properties are shown in Table 1.\\
	\begin{table}[b]
		\begin{center}
  		\caption{Ring-linear DNA blends studied in this work. The total DNA concentration in all blends was maintained at 50 $\mu$g/mL.}
 		 \label{tbl:example}
 		 \begin{tabular}{cccccc}
   		 \hline
   			 	  Ring DNA & Ring DNA  & Linear DNA & Linear DNA  &  Volume\\
   			 		($\%$)   & mass ($\mu$g) & ($\%$)  & mass ($\mu$g) &  (mL) \\
   		\hline
    				 0 & 0 & 100 & 250 &  5\\
    				17 & 42.5 & 83 & 207.5 &  5\\
    				50 & 125 & 50 & 125 &  5\\
    				83 & 207.5 & 17 & 42.5 &  5\\
    	\hline
 		 \end{tabular}
  		\end{center}
	\end{table}
	
\textbf{C. Fluorescent labeling of ring DNA}

	For single molecule imaging, DNA is fluorescently labeled with an intercalating dye (YOYO-1, Molecular Probes, Thermo Fisher) at a dye-to-base pair ratio of 1:4 for $>$1 hour in the dark at room temperature. Trace amounts of fluorescently labeled ring DNA are then added to background solutions of unlabeled semidilute ring-linear DNA blends, resulting in a final labeled DNA concentration of $2\times10^{-3}$  $\mu$g/mL. In addition, a small amount of the reducing agent $\beta$-mercaptoethanol (14 $\mu$M) and coupled enzymatic oxygen scavenging system containing glucose (50 $\mu$g/mL), glucose oxidase (0.01 $\mu$g/mL), and catalase (0.004 $\mu$g/mL) are added into the ring-linear DNA blends to suppress photobleaching and photocleaving of fluorescently labeled DNA molecules. The semidilute ring-linear blend mixture is rotationally mixed for $>$20 minutes before single molecule imaging. Solution viscosity $\eta_\mathrm{s}$ is determined using a cone and plate viscometer (Brookfield, USA) at 22.5 $^\circ$C.\\
	
\textbf{D. Optics and imaging}

	Single molecule fluorescence microscopy is performed using an inverted epifluorescence microscope (IX71, Olympus) coupled to an electron-multiplying charge coupled device (EMCCD) camera (iXon, Andor Technology). Fluorescently labeled DNA samples are illuminated using a 50 mW 488 nm laser (Spectra-Physics, CA, USA) directed through a neutral density (N.D.) filter (ThorLabs, NJ, USA) and a 488 nm single-edge dichroic mirror (ZT488rdc, Chroma). Fluorescence emission is collected by a 1.45 NA, 100$\times$ oil immersion objective lens (UPlanSApo, Olympus) followed by a 525 nm single-band bandpass filter (FF03-525/50-25, Semrock) and a 1.6$\times$ magnification lens, yielding a total magnification of 160$\times$. Images are acquired by an Andor iXon EMCCD camera (512$\times$512 pixels, 16 $\mu$m pixel size) under frame transfer mode at a frame rate of 33 Hz (0.030 s$^{-1}$). Images obtained using fluorescence microscopy are analyzed using a custom \textsc{Matlab} code to quantify the polymer conformations, as previously described \cite{Zhou2019}. The full contour length of fluorescently labeled 45 kbp ring DNA is approximately 20 $\mu$m, such that the stretched contour length of the 45 kbp ring polymer is $L_\mathrm{{circ}}$ = 10 $\mu$m \cite{Zhou2019}, which is equal to one-half of the fully stretched contour length of the equivalent linear polymer of identical molecular weight, $L_\mathrm{lin}$.\\

\textbf{E. Microfluidics and flow field characterization}

	Two-layer PDMS microfluidic devices are fabricated using standard techniques in soft lithography, as previously described \cite{Zhou2016b}. In brief, the microfluidic device contains a fluidic layer situated below a control layer containing a fluidic valve. The fluidic layer is fabricated to contain a cross-slot channel geometry with 300 $\mu$m in width and 100 $\mu$m in height. In this way, planar extensional flow is generated in the fluidic layer, and the control layer contains a pressure-driven valve to control fluid flow. Flow field characterization and strain rate determination are performed in ring-linear blends prior to single polymer dynamics experiments using particle tracking velocimetry (PTV), as previously described \cite{Zhou2016}.\\

\textbf{F. Modeling and simulation of semidilute ring-linear blends}

Ring-linear polymer solution blends are modeled by coarse-grained bead-spring chains with approximately 1 Kuhn step per spring. We perform BD simulations to evolve bead positions in an implicit solvent via the Langevin equation of motion:
\begin{equation}
    \frac{d\tilde{\bm{r}}_{i}}{d\tilde{t}} = \tilde{\bm{\kappa}} \cdot \tilde{\bm{r}}_{i} -\sum_{j} \tilde{\textbf{D}}_{ij} \nabla_{\tilde{\bm{r}}_{j}}(\tilde{U}) + \tilde{\bm{\xi}}_{i}
\end{equation}
where tildes denote dimensionless quantities.  Polymer beads experience flow via the $3N \times 3N$ block diagonal tensor $\tilde{\bm{\kappa}}$, which has $3 \times 3$ diagonal blocks given by the solvent velocity gradient tensor $(\nabla \tilde{\textbf{v}})^T$. Conservative interactions $\tilde{U}$ are given by the Kremer-Grest potential, which accounts for finite extensibility and prevents spring crossings \cite{kremer1990dynamics}. Solvent-mediated HI and Stokes drag are included via the diffusion tensor $\tilde{\textbf{D}}_{ij}$, for which we use the Rotne-Prager-Yamakawa tensor \cite{rotne1969variational, yamakawa1970transport}. The Brownian noise $\tilde{\bm{\xi}}_i$ is approximated by the truncated expansion ansatz \cite{geyer2009n}. Evaluation of the diffusion tensor and Brownian noise is accelerated by the iterative conformational averaging method \cite{miao2017iterative, young2018conformationally, young2019simulation}. We simulate ring and linear polymers with an equal number of beads per chain $N_R = N_L = 150$. Generally we consider $N_C = 128$ chains per simulation, with the number of rings and linear chains chosen to match the blend ratios in experiments. Further details and verification of the method are available in previous work \cite{young2018conformationally, young2019simulation}, and generalization to ring-linear polymer blends is described in a companion article \cite{young2020dynamics}.

\section{Results and Discussion}
\textbf{A. Longest relaxation time of ring polymers in ring-linear blends}
		 \begin{figure*}[t]
 			\includegraphics[scale=1.1]{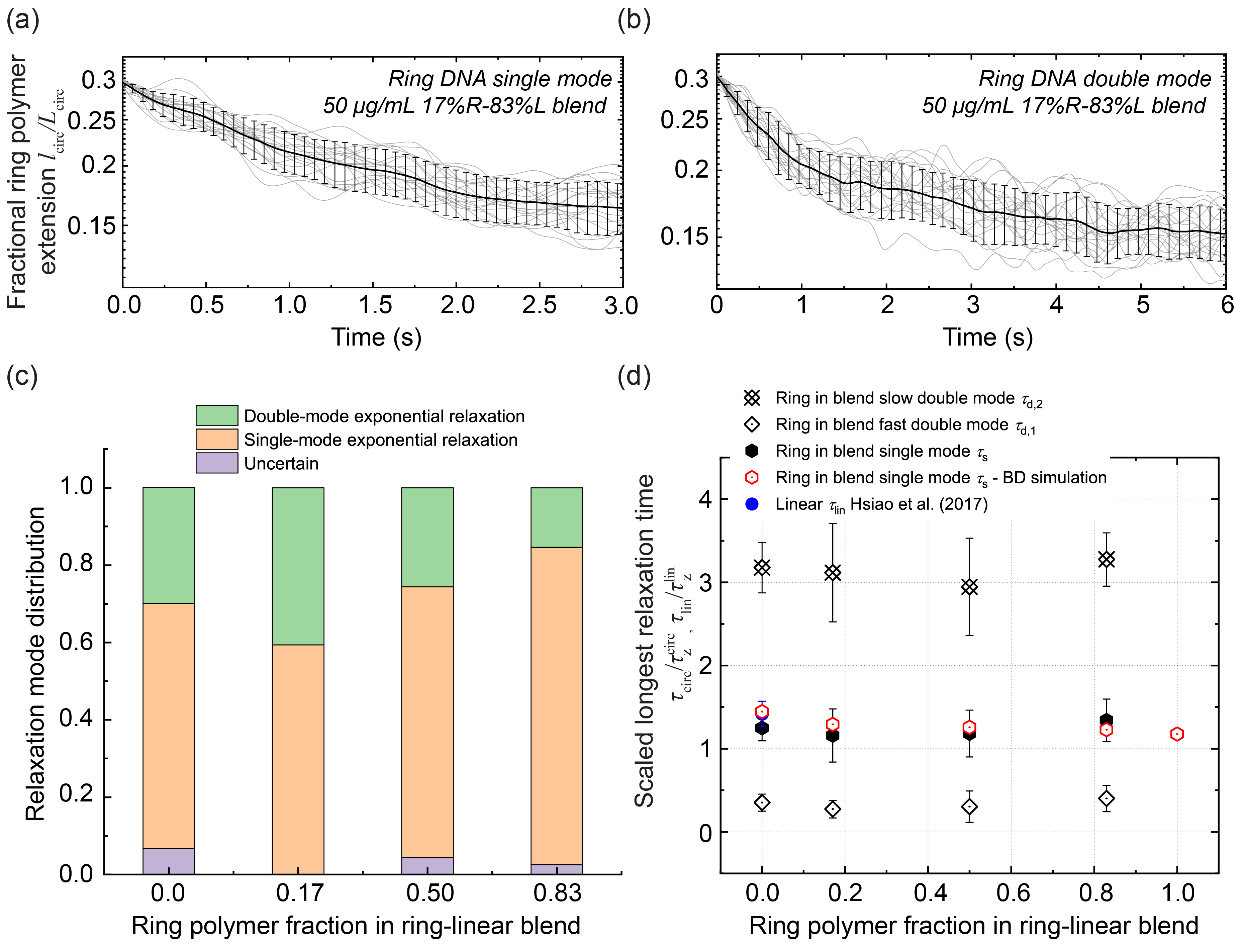}
  			\caption{Relaxation of ring polymers in semidilute ring-linear polymer blends. Single molecule relaxation trajectories (grey) and ensemble-averaged relaxation trajectories (black) for molecular sub-populations corresponding to (a) single-mode and (b) double-mode exponential relaxation trajectories for ring polymers in a semidilute background solution of 17$\%$ ring - 83$\%$ linear blend at 50 $\mu$g/mL. Error bars are determined from standard deviation of molecular trajectories. (c) Fraction of single-mode and double-mode exponential relaxation behavior as a function of ring polymer fraction in blends. Uncertainties in classifying relaxation behavior are also plotted for cases where single molecule trajectories are not well described by single-mode or double-mode behavior. (d) Longest relaxation times normalized by dilute solution values $\tau_z^{circ}$ for ring polymers and $\tau_z^{lin}$ for linear polymers from experiments (back diamonds) and BD simulations (red hexagons) in semidilute ring-linear polymer blends as a function of ring polymer fraction. Longest relaxation time for linear polymers in semidilute unentangled solution is also shown as a reference (blue circle) \cite{Hsiao2017}. Experimental molecular ensembles consist $n \geq 50$ single molecules for each blend.}
  			\label{Relaxation}
		\end{figure*}

	We began by determining the longest relaxation time of ring polymers in semidilute ring-linear blends with varying ring fraction (Fig. \ref{Relaxation}). In all cases, the total polymer concentration was maintained at 50 $\mu$g/mL, which corresponds to the overlap concentration, $c^*$, of linear 45 kbp DNA. Ring-linear blends are subjected to a step strain in planar extensional flow at a strain rate $\dot{\epsilon}$ above the coil-stretch transition. The accumulated fluid strain or Hencky strain, $\epsilon$, is calculated as $\epsilon = \int_0^{t_d} \dot{\epsilon}(t') dt'$, where $t_d$ is the duration of the step strain rate input, $\dot{\epsilon}(t') = \dot{\epsilon} H(t')$, and $H$ is the Heaviside function. Following the step strain rate input (at times $t>t_d$), the flow is stopped and the polymer solution relaxes back to equilibrium. In this way, single polymer relaxation trajectories are obtained as part of the step strain-relaxation experiments, such that fluorescently labeled ring DNA molecules experience at least $\epsilon$ = 20 units of fluid strain prior to cessation of extensional flow. During the flow portion of this step, fluorescently labeled rings are stretched to $\approx$0.6$ L_\mathrm{circ}$ prior to flow cessation. Longest polymer relaxation times are then determined by fitting the terminal 30$\%$ average squared fraction extension $(l_\mathrm{circ}/L_\mathrm{circ})^2$ to a single-mode or double-mode exponential decay function, as previously described \cite{Zhou2018,Schroeder2018,Zhou2019} and elaborated on below. Here $l_\mathrm{circ}$ denotes the experimentally measured span of polymer extension in the two-dimensional flow plane.
	
	Our results reveal two distinct molecular sub-populations for ring polymer relaxation in semidilute ring-linear blend solutions. One molecular sub-population relaxes via a single exponential decay, whereas the second molecular sub-population relaxes via a double exponential decay response. For instance, ring polymer relaxation in a 17$\%$ R-83$\%$ L blend shows that approximately 60$\%$ of the molecular relaxation trajectories are well described by a single-mode exponential decay (Fig. \ref{Relaxation}a), whereas approximately 40$\%$ of the relaxation trajectories are found to exhibit double-mode exponential decay (Fig. \ref{Relaxation}b). Here, the single-mode relaxation time $\tau_s$ is determined from $(l_\mathrm{circ}/L_\mathrm{circ})^2 =A \exp(-t/ \tau_s) + B$, where $A$ and $B$ are numerical constants. The fast and slow double-mode relaxation times $\tau_{d,1}$ and $\tau_{d,2}$ are determined from $(l_\mathrm{circ}/L_\mathrm{circ})^2 = A_1 \exp(-t/ \tau_{d,1}) + A_2 \exp(-t/ \tau_{d,2}) +B$, where $A_1$, $A_2$, and $B$ are numerical constants. In order to distinguish between single-mode and double-mode exponential decay behavior, molecular relaxation trajectories are fit to both functions to determine the best fit behavior, as previously described \cite{Zhou2019}.   
	
	Interestingly, the fraction of single-mode versus double-mode relaxation trajectories depends on ring-linear polymer blend composition (Fig. \ref{Relaxation}c). In particular, the fraction of double-mode relaxation trajectories shows a maximum for ring-linear blends with 17$\%$ R-83$\%$ L composition and decreases upon further increases in ring fraction. A similar trend of heterogeneous relaxation of ring polymers was recently observed in ring-linear blend melts using MD simulations \cite{Katsarou2020}, where the distribution width of different relaxation modes for rings decreased with increasing ring composition in the blend. However, it is important to note that the heterogeneous relaxation behavior for rings observed here is qualitatively different than the relaxation of linear polymers in ultra-dilute solutions (10$^{-5}$ $c^*$) \cite{Schroeder2018}, linear polymers in semidilute solutions of purely linear chains \cite{Hsiao2017, Zhou2018}, and ring polymers in ultra-dilute solutions (10$^{-5}$ $c^*$) \cite{Li2015, Hsiao2016a}, all of which exhibit a simple single-mode exponential decay for polymer relaxation in the terminal regime. In semidilute unentangled solutions of purely linear chains (0$\%$ R-100$\%$ L), ring polymers similarly exhibit two distinct molecular sub-populations showing single and double-mode relaxation behavior \cite{Zhou2019}. 
	
	The bimodal relaxation behavior is thought to arise from transient intermolecular structures wherein linear polymer chains thread into open ring polymers. Here, ring relaxation is influenced by the presence of threaded linear chains, resulting in a large fraction of double-mode relaxation trajectories for blends with low to intermediate ring fractions (e.g. blend composition of 17$\%$ R-83$\%$ L). We posit that semidilute ring-linear blends with low ring fractions provide a diverse set of local environments in terms of intermolecular HI between rings and linear chains and concentration fluctuations that qualitatively differs from purely linear semidilute polymer solutions, thereby giving rise to the double-mode relaxation behavior. Upon further increasing the ring fraction in ring-linear blends, threading interactions between rings and linear polymers become increasingly less likely such that, on average, rings tend to relax in the absence of intermolecular threading interactions. In addition, the slow phase of the double-mode ring relaxation can also be influenced by solvent-mediated hydrodynamic coupling to linear polymer chain relaxation, discussed in detail in the companion article \cite{young2020dynamics}, which is supported by the observation that $\tau_z^{lin} \approx 3 \tau_z^{circ} \approx \tau_{d,2}$, where $\tau_z$ denotes the dilute solution longest relaxation time. Indeed, local concentration fluctuations in lightly entangled polymer solutions are known to give rise to multiple molecular sub-populations governing polymer relaxation \cite{Zhou2018}. 
	
	Longest relaxation times of ring polymers are plotted in Fig. \ref{Relaxation}d as a function of ring fraction in ring-linear blends. The longest relaxation time for linear polymers in semidilute solutions (1 $c^*$) of purely linear polymers is also plotted as a reference \cite{Hsiao2017}. Moreover, ring polymer relaxation times are included from BD simulations. Relaxation times for ring polymers, including the single-mode timescale $\tau_s$ and double-mode timescales $\tau_{d,1}$ and $\tau_{d,2}$, and the longest relaxation time for linear polymers, $\tau_{lin}$, are normalized by their corresponding longest relaxation times in the dilute limit, denoted as $\tau_z^{circ}$ and $\tau_z^{lin}$, respectively. Non-normalized quantities for all relaxation times are tabulated in Supplementary Table 1. 
	
	Single- and double-mode relaxation times are relatively independent of ring-linear blend compositions at a total polymer concentration of 50 $\mu$g/mL, or 1 $c^*$ corresponding to the linear DNA polymer. The normalized single-mode relaxation time for ring polymers $\tau_s$ in ring-linear blends is consistent with the relaxation time for pure linear polymers in semidilute solutions, which supports the hypothesis that the single-mode exponential relaxation behavior corresponds to relaxation of polymers free from topological interactions with surrounding molecules, regardless of chain topology. The normalized slower double-mode relaxation time $\tau_{d,2}$ is approximately 3$\times$ larger than the normalized single-mode relaxation time, whereas the faster double-mode relaxation time $\tau_{d,1}$ is nearly 10$\times$ smaller than the normalized single-mode relaxation time. Interestingly, the single- and double-mode relaxation times appear to be insensitive to blend compositions. Therefore, similar to ring polymer relaxation in pure semidilute linear polymer solutions \cite{Zhou2019}, we posit that double-mode relaxation behavior for rings in ring-linear blends originates from the formation of transient threaded structures between rings and linear polymers. Moreover, our results suggest that these structures are local and do not include long-range linked intermolecular structures in semidilute solutions. However, the probability of forming ring-linear transient structures varies with blend composition, which is reflected in the different fractions of single- and double-mode relaxation behavior in Fig. \ref{Relaxation}c. Interestingly, prior diffusion measurements show that ring DNA polymer diffusion coefficients remain relatively constant in 100 $\mu$g/mL blends with different ring-linear compositions \cite{Chapman2012a}, which is consistent with our results for longest relaxation time.
		
	Overall, our results show clear differences between the ring relaxation behavior in semidilute pure linear polymer solutions and ring-linear polymer blends. For semidilute unentangled solutions of purely linear polymers ($c \approx c^*$), strictly single-mode relaxation behavior for linear chains is observed \cite{Hsiao2017}. Interestingly, for solutions of purely linear polymers, double-mode relaxation behavior begins to emerge only when the solution concentration is above the critical entanglement concentration $c_e$ such that $c>c_e$ \cite{Zhou2018}. On the other hand, our results for ring-linear blends show two distinct molecular sub-populations for ring relaxation in semidilute unentangled solutions at concentrations $c \approx c^*$. Taken together, these results highlight the importance of molecular topology on the relaxation dynamics of ring polymers in different ring-linear polymer blend solutions.\\
	
	\textbf{B. Transient stretching dynamics of rings in ring-linear blends}
		 \begin{figure*}
 			\includegraphics[scale=1]{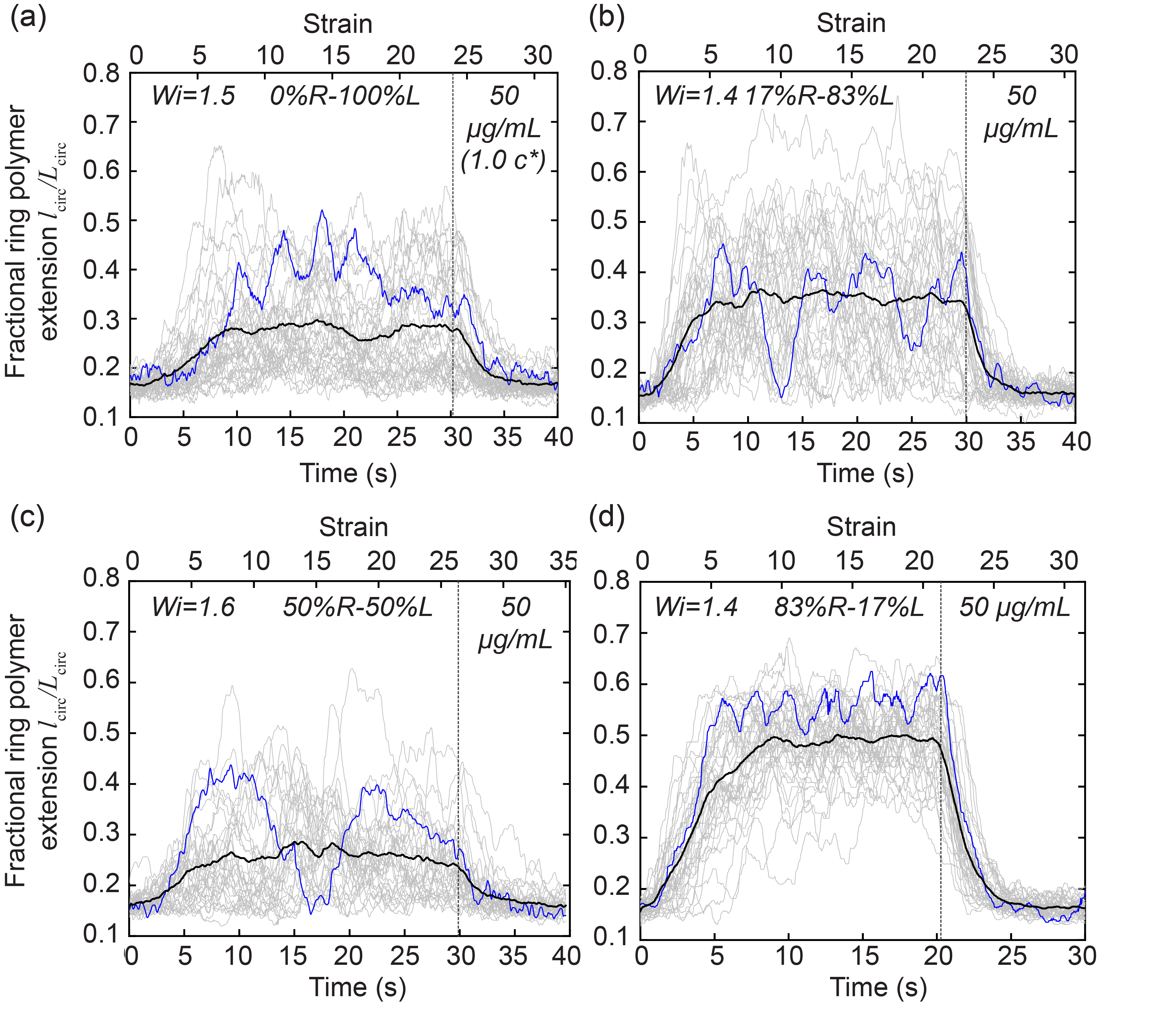}
  			\caption{Single molecule trajectories of ring polymers in semidilute ring-linear polymer blends show large conformational fluctuations. Transient fractional extension of ring DNA polymers in 50 $\mu$g/mL semidilute ring-linear polymer blends at $Wi \approx$ 1.5 with (a) 0$\%$ ring - 100$\%$ linear polymers (b) 17$\%$ ring - 83$\%$ linear polymers, (c) 50$\%$ ring - 50$\%$ linear polymers, and (d) 83$\%$ ring - 17$\%$ linear polymers. Individual single molecule trajectories are shown in gray lines and ensemble averaged trajectories are shown in a black line. A characteristic individual single molecule trajectory is highlighted in blue line. Molecular ensembles consist of $n = 38$, $n = 40$, $n = 34$, and $n = 39$ molecules for ring polymers in 0$\%$ R-100$\%$ L, 17$\%$ R-83$\%$ L, 50$\%$ R-50$\%$ L, and 83$\%$ R-17$\%$ L blends, respectively. The dashed line indicates the time at which the step strain input is stopped.}
  			\label{SM_traj}
		\end{figure*}
	
	We next investigated the transient stretching dynamics of ring polymers in semidilute ring-linear blends with different ring fractions. In all cases, the total polymer concentration was maintained at 50 $\mu$g/mL. In these experiments, fluorescently labeled rings are first allowed to relax to an equilibrium conformation for at least 2$\tau_s$ in the absence of flow. At time $t=0$, a step strain rate input with precisely controlled strain rate $\dot{\epsilon}$ is imposed on the polymer blend sample for total fluid strain $\epsilon$. The flow strength is characterized by the Weissenberg number $Wi = \dot{\epsilon} \tau_s$, which is defined by the strain rate nondimensionalized by the single-mode relaxation time $\tau_s$. During the stretching phase, a single fluorescently labeled ring polymer is confined near the stagnation point of planar extensional flow using automated flow control in a device known as Stokes trap \cite{Shenoy2016, Zhou2016}. In this way, ring polymers are trapped for long residence times in extensional flow with well-defined strain rates, enabling direct observation of transient and steady stretching dynamics. During this process, only minor corrections are made to the inlet pressure for flow control, such that the strain rate remains constant during the flow phase of the experiment \cite{Shenoy2016, Zhou2016}. Following the step deformation, the flow is abruptly halted (denoted by the dashed line in Fig. \ref{SM_traj}), and ring polymers are allowed to relax back to the thermal equilibrium, as discussed in Sec III.A.
	
	Fig. \ref{SM_traj} shows the transient fractional extension $l_\mathrm{circ}/L_\mathrm{circ}$ of ring polymers in semidilute ring-linear polymer blend near $Wi \approx 1.5$ for four different blend compositions. Additional results for $Wi = 1$ and $Wi = 2.5$ are shown in Supplementary Fig. 1. In all cases, ring polymers are subjected to  $\epsilon > 20$ units of fluid strain, and 30-40 single molecule trajectories are analyzed for each condition. In Fig. \ref{SM_traj}, the black curves represent the ensemble-averaged fractional extension over all single molecule trajectories. Individual single molecule trajectories are plotted in gray, and one representative single molecule trajectory is highlighted in blue. 
	
	Transient stretching trajectories show that ring polymers exhibit large fluctuations in chain extension in extensional flow. Interestingly, conformational fluctuations are observed for ring polymers in all ring-linear blends. Chain fluctuations persist long after the initial transient stretching process has ended, such that chain fluctuations continue even for large amounts of accumulated strain $\epsilon >$ 8-10. Qualitatively, rings exhibit large conformational fluctuations in ring-linear blends with intermediate ring fractions, such as the 17$\%$ R-83$\%$ L blend. Moreover, conformational fluctuations appear to decrease upon further increasing the ring fraction towards the 83$\%$ R-17$\%$ L blend. These trends are consistent with the fraction of single- and double-mode relaxation trajectories in ring-linear blends, as shown in Fig. \ref{Relaxation}c. 
	
	BD simulations show that ring polymer conformational fluctuations are driven by {\em both} intermolecular threading events with nearby linear chains and solvent-mediated intermolecular HI \cite{young2020dynamics}. These behaviors are discussed in detail in the companion simulation and modeling article \cite{young2020dynamics}, and we briefly summarize these findings here with respect to experimental results. In general, highly stretched linear chains in semidilute ring-linear blend solutions induce strong hydrodynamic disturbance flows, generally stronger than more compact ring polymers at the same flow strength. Consequently, these long-range HI disturbance flows drive large conformational fluctuations in ring polymers in flow. 
	
	In single molecule experiments, we observe unexpected polymer stretching behavior in semidilute solutions that is attributed to intermolecular HI. For example, for 17$\%$ R-83$\%$ L and 50$\%$ R-50$\%$ L blends, some ring polymers fully recoil back to equilibrium levels of extension, followed by re-stretching during continued deformation in flow (blue trajectory in Fig. \ref{SM_traj}c). This behavior resembles the large conformational fluctuations that arise due to polymer tumbling in dilute solution shear flow \cite{Smith1999}, although with completely different physical origins. In the case of dilute solution shear flow, the coupling between the rotational and extensional components of flow leads to tumbling behavior for rings \cite{Tu2020} and linear polymers \cite{Smith1999,Schroeder2018}. For semidilute ring-linear blends in extensional flow, however, conformational fluctuations arise due to flow-driven intermolecular interactions and long-range HI between polymers with different chain topologies and relaxation times. Upon further increasing the ring fraction, such as in the 83$\%$ R-17$\%$ L blend, ring polymers eventually show smaller magnitude conformational fluctuations and generally stretch to larger fractional extensions. This behavior arises due to a decreased probability of ring-linear threading interactions and decreased magnitude HI disturbance flows in blends with dominant ring fraction, as stretched linear chains generally induce stronger intermolecular HI disturbance flows in ring-linear blends \cite{young2020dynamics}. 
	
		 \begin{figure*}
 			\includegraphics[scale=1.6]{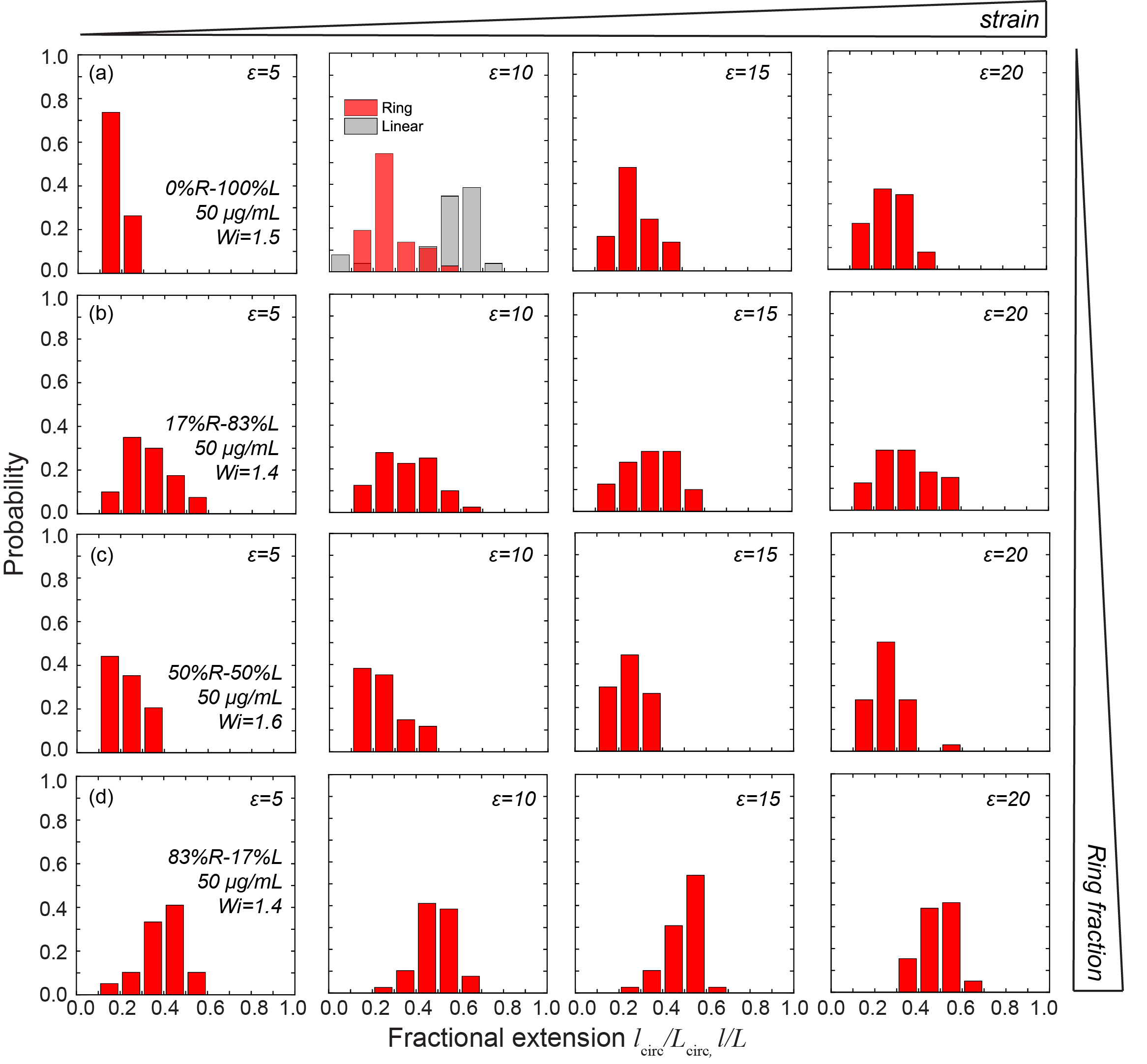}
  			\caption{Probability distribution of ring polymer fractional extension in semidilute ring-linear blends as a function of blend composition near $Wi \approx$ 1.5. Histograms showing probability of ring polymer extension in semidilute ring-linear polymer blends with: (a) 0$\%$ ring -100$\%$ linear, (b) 17$\%$ ring - 83$\%$ linear, (c) 50$\%$ ring - 50$\%$ linear, and (d) 83$\%$ ring - 17$\%$ ring for accumulated fluid strains of $\epsilon=5$, $\epsilon=10$, $\epsilon=15$, $\epsilon=20$ at $Wi \approx$ 1.5. Molecular ensembles consist of $n = 38$, $n = 40$, $n = 34$, and $n = 39$ molecules for four different blends, respectively. All experiments are performed at a total polymer concentration of 50 $\mu$g/mL. Data for linear polymers (gray bars) in semidilute solutions of pure linear chains at 50 $\mu$g/mL are from \citet{Hsiao2017}.}
  			\label{Probability}
		\end{figure*}
	
	The large conformational fluctuations for ring polymers in semidilute ring-linear blends gives rise to broad distributions in ring polymer extension in flow. Histograms of ring polymer extension for four different blend compositions near $Wi \approx$ 1.5 are shown in Fig. \ref{Probability}. Compared to linear polymers in pure semidilute linear polymer solutions, ring polymers in pure semidilute linear polymer solutions (0$\%$ R-100$\%$ L) exhibit broader distributions of molecular extension at smaller fractional extensions (Fig. \ref{Probability}a). For instance, linear polymers are stretched to $l_\mathrm{lin}/L >$ 0.6 at $\epsilon$ = 10, while the majority of ring polymers shows a fractional extension of $l_\mathrm{circ}/L_\mathrm{circ} \approx 0.3$. The probability distribution of ring polymer extension $l_\mathrm{circ}/L_\mathrm{circ}$ further broadens in the 17$\%$ R-83$\%$ L blend (Fig. \ref{Probability}b). Interestingly, ring polymer extension $l_\mathrm{circ}/L_\mathrm{circ}$ ranges from $\approx$ 0.2-0.6 and is peaked around $l_\mathrm{circ}/L_\mathrm{circ} \approx 0.4$ when the accumulated fluid strain increases from $\epsilon$ = 5 to 20. This behavior is consistent with the single molecule transient stretching trajectories for the 17$\%$ R-83$\%$ L blend (Fig. \ref{SM_traj}b). Moreover, the broad distribution of $l_\mathrm{circ}/L_\mathrm{circ}$ is consistent with the notion that large conformational fluctuations result from intermolecular interactions between rings and linear polymers in the blend. Upon further increasing the ring fraction in the blend to 50$\%$ R-50$\%$ L, the probability distribution of $l_\mathrm{circ}/L_\mathrm{circ}$ for rings narrows and shifts toward a smaller extension, with the majority of ring polymers stretched to $l_\mathrm{circ}/L_\mathrm{circ} \approx 0.3$ (Fig. \ref{Probability}c). These results are consistent with observations from the single molecule trajectories (Fig. \ref{SM_traj}c), where the 50$\%$ R-50$\%$ L blend shows the smallest average fractional extension. Upon increasing the ring fraction further to 83$\%$ R-17$\%$ L (Fig. \ref{Probability}d), the distributions shift to larger average extensions and rings tend to become more stretched for larger fluid strains. \\
	
	\textbf{C. Molecular individualism and ring polymer conformational fluctuations}
		 \begin{figure*}
 			\includegraphics[scale=1.1]{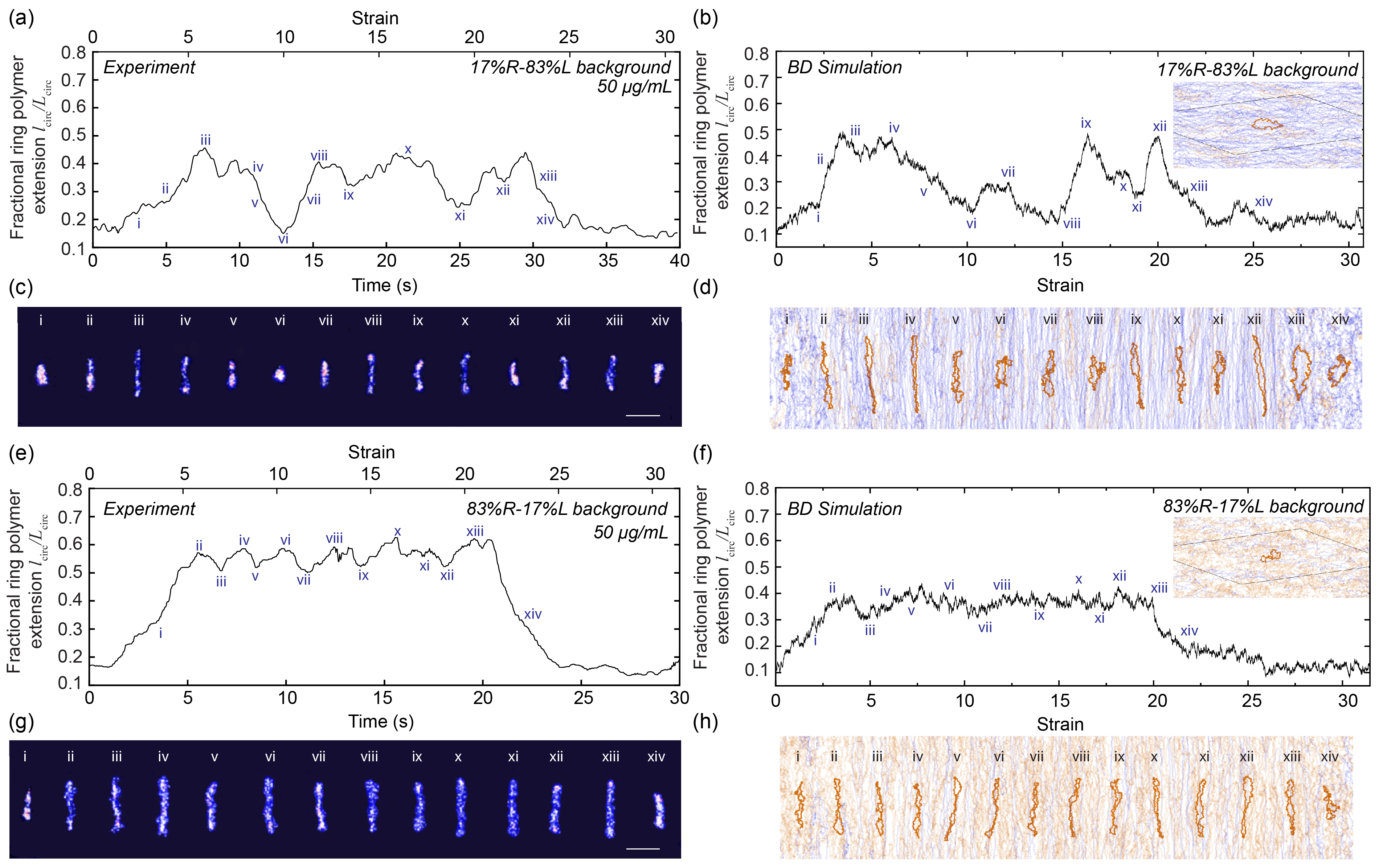}
  		\caption{Characteristic transient stretching trajectories for ring polymers in semidilute 17$\%$ ring - 83$\%$ linear blends at $Wi \approx 1.5$ from (a) experiments and (b) BD simulations. Corresponding single molecule snapshots are shown in (c) and simulation snapshots are shown in (d). Characteristic transient stretching trajectories for ring polymers in semidilute 83$\%$ ring - 17$\%$ linear blends at $Wi \approx 1.5$ from (e) experiments and (f) BD simulations. Corresponding single molecule snapshots are shown in (g) and simulation snapshots are shown in (h). The Roman numerals correspond to individual time points along the trajectory. The scale bar is 3 $\mu$m in the experimental snapshots. In the simulation snapshots, linear polymers are labeled in blue and ring polymers are labeled in yellow.}
			\label{SMsnapshots}
		\end{figure*}
	
	Characteristic transient trajectories for single ring polymers from single molecule experiments and BD simulations are shown in Fig. \ref{SMsnapshots}. A characteristic stretching trajectory for a ring polymer in a 17$\%$ R-83$\%$ L blend at $Wi$ $\approx$ 1.5 is shown in Fig. \ref{SMsnapshots}a, together with corresponding single polymer snapshots during the stretching trajectory (Fig. \ref{SMsnapshots}c). The Roman numerals correspond to individual time points along the trajectory. Large conformational fluctuations are observed for ring polymers in 17$\%$ R-83$\%$ L blends, with some rings unexpectedly showing recoiling and subsequent restretching behavior in flow, as shown in Fig. \ref{SMsnapshots}a (denoted by time point vi). In simple shear flow, ring polymers exhibit repeated cycles of stretching, collapse, and tumbling which arises due to a coupling between the rotational and extensional components of flow. Although extensional flow alone is inadequate to produce the characteristic tumbling behavior observed in shear flow, semidilute blends of ring-linear chains facilitate long-range HI that can generate local disturbance flows. Together with intermolecular threading interactions, these effects give rise to large-scale conformational fluctuations in flow. 
	
	We further visualized ring conformational fluctuations with molecular simulation, and we show a characteristic dynamic trajectory for a single ring polymer in a 17$\%$ R-83$\%$ L blend at $Wi$ = 1.5 in Fig. \ref{SMsnapshots}b. We again plot the fractional ring polymer extension $l_{\text{circ}}/L_{\text{circ}}$ as a function of strain, denoting snapshots (Fig. \ref{SMsnapshots}d) at individual time points along the trajectory with Roman numerals. In agreement with the single molecule experiments, BD simulations show large magnitude conformational fluctuations for rings in flow, including large retraction and restretching events in semidilute solution extensional flow. We note that this behavior highlights the importance of long-range HI in generating local flows that are non-extension dominated and contain rotational character, in part because the ring polymer shown in this characteristic trajectory does not exhibit a distinct ring-linear threading event that drives conformational fluctutations. We further explore the role of HI in driving ring conformational fluctuations in the companion article \cite{young2020dynamics}.
	
	Prior work has reported that ring conformational fluctuations in semidilute linear polymer solutions can also be caused by threading of linear polymers into the partially stretched, open conformation of rings \cite{Zhou2019}. Transport of linear polymer chains into open ring polymers in flow leads to repeated threading and unthreading events, giving rise to the repeated cycles of transient chain extension and retraction. Upon increasing the fraction of rings in ring-linear blends, threaded linear chains can (in  principle) simultaneously interact with adjacent rings in the blend solution, thereby giving rise to complex intermolecular interactions. Our results suggest that large magnitude conformational fluctuations for rings arise not only due to long-range HI, but also due to the formation of transient ring-linear threaded structures in the blend. Experimental SMFM results show that rings fluctuate drastically in non-dilute flows, and in some cases fully recoil (Figs. \ref{SMsnapshots}a-d) in low to intermediate ring fraction blends (17$\%$R-83$\%$L). Interestingly, ring conformational fluctuations drastically decrease as the ring fraction increases (e.g. 83$\%$ R-17$\%$ L blends), showing only minor fluctuations in ring polymer fractional extension (Fig. \ref{SMsnapshots}e,g). Analogous results are  observed in 83$\%$ R-17$\%$ L blends in BD simulations (Figs. \ref{SMsnapshots}f,h), which similarly show smaller magnitude conformational fluctuations at larger ring fraction. These observations are consistent with the notion of a decreased probability of ring-linear interactions with the majority of ring polymers in the blend, including both intermolecular threading events and solvent-mediated hydrodynamic interactions. 
	
	We quantify ring conformational fluctuations by determining the average fluctuation in fractional chain extension, $\langle \delta \rangle / L_\mathrm{circ}$. Here, $\langle \delta \rangle / L_\mathrm{circ}$ is defined as the fluctuation quantity in  conformational extension fluctuation over the molecular ensemble after the initial transient stretching phase such that:
	\begin{equation}
		\frac{\langle \delta \rangle}{L_\mathrm{{circ}}} = \frac{\sum_{n=1}^{\mathcal{N}}\sqrt{ \sum_{t_{90}}^{t_f} [l_n(t) - \langle l_n \rangle]^2}}{\mathcal{N} L_\mathrm{{circ}}}
		\label{fluc_quant}
	\end{equation}	
	where $l_n(t)$ is the transient polymer extension for individual single polymer chains at time $t$, $\langle l_n \rangle$ is the time-averaged or mean polymer extension, and $\mathcal{N}$ denotes the total number of individual trajectories in the ensemble. In Eq. \ref{fluc_quant}, $t_\mathrm{f}$ denotes the time when the step strain rate input stops (dashed line in Fig. \ref{SM_traj}), and $t_{90}$ is defined as the time at which the fractional polymer extension first reaches 90$\%$ of the average fractional extension at time $t_\mathrm{f}$. In this way, we discard the initial transient stretching of ring polymers and only compute the chain extension fluctuation quantities after the initial transient phase has died out, as previously reported \cite{Zhou2019}. In all cases, transient polymer extension is observed for at least 20-25 units of strain in extensional flow.
	
		 \begin{figure}
 			\includegraphics[scale=1.1]{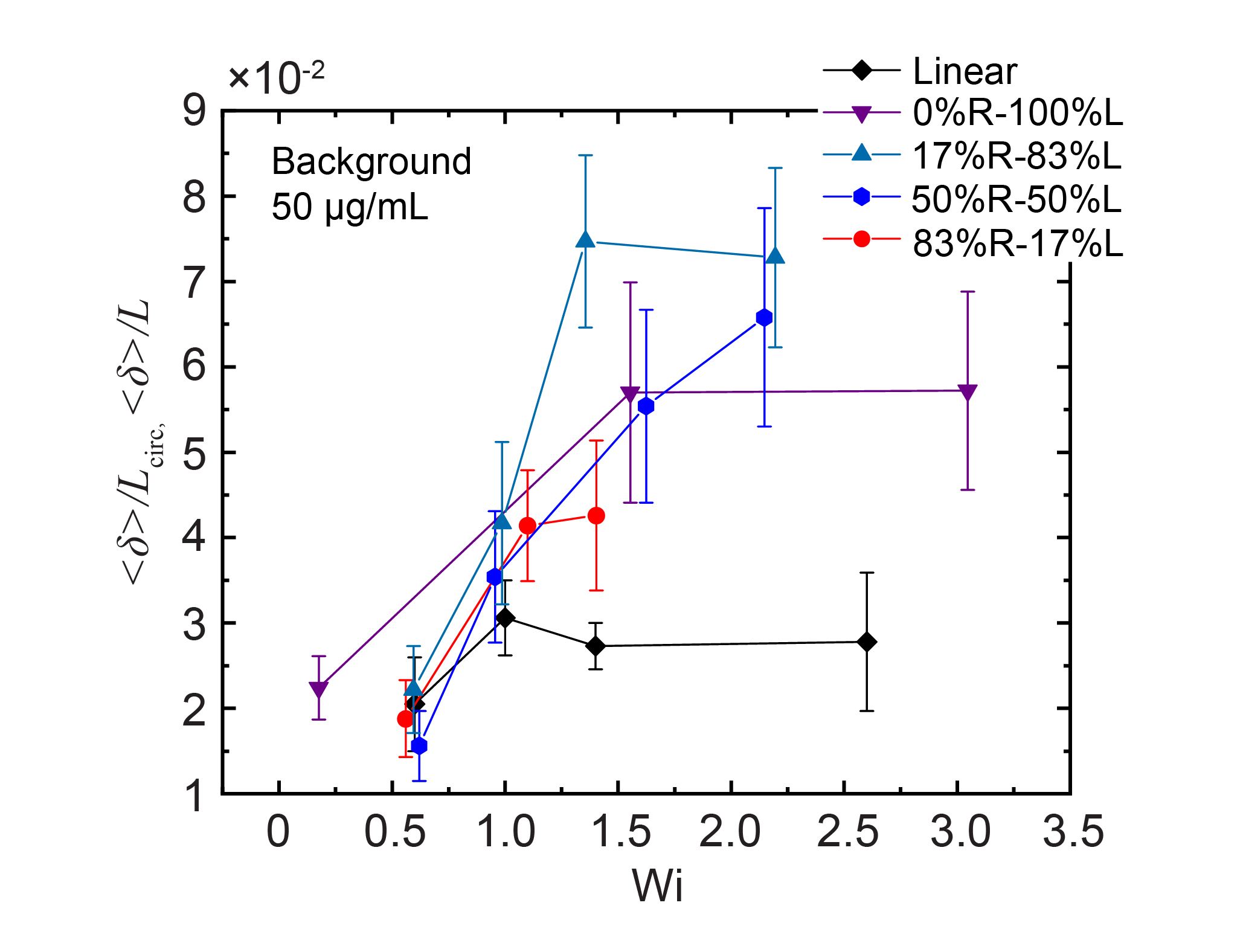}
  			\caption{Fluctuations in fractional extension of ring polymers in semidilute ring-linear polymer blends as a function of $Wi$. The average chain fluctuation quantity $\langle \delta \rangle$ is normalized in terms of the contour length $L_\mathrm{circ}$ for ring polymers and $L$ for linear polymers, respectively. Each molecular ensemble contains $n \geq 34$ single molecule traces.}
  			\label{conf_fluc}
		\end{figure}
	As shown in Fig. \ref{conf_fluc}, fractional extension fluctuations for ring polymers in semidilute ring-linear blends $\langle \delta \rangle / L_{\mathrm{circ}}$ are compared to linear polymer chain fluctuations in pure semidilute linear polymer solutions $\langle \delta \rangle / L$. Our results show an increase in ring chain fluctuations in semidilute ring-linear polymer blends compared to pure linear polymer solutions. Moreover, ring polymer chain fluctuations in ring-linear blends do not increase monotonically with increasing ring fraction. Ring fluctuations first increase upon increasing the ring fraction in blends, reaching a maximum for 17$\%$ R-83$\%$ L blends, followed by a gradual decrease in magnitude for the 83$\%$ R-17$\%$ L blends. This behavior is consistent with trends observed in the probability distribution of relaxation modes (Fig. \ref{Relaxation}b) and for single molecule trajectories (Fig. \ref{SM_traj}) showing a unique set of molecular individualism (Fig. \ref{SMsnapshots}), wherein ring conformations show large fluctuations including full retraction to a coiled state during the course of strong extensional flow in 17$\%$ R-83$\%$ L and 50$\%$ R-50$\%$ L blends. Upon increasing the fraction of rings in ring-linear blends, intermolecular interactions between stretched rings and linear chains are gradually suppressed, which results in a eventual decrease in chain fluctuations. In addition, ring polymer chain fluctuations tend to increase in the vicinity of the coil stretch transition (CST), corresponding to Weissenberg numbers near $Wi \approx 1.0$ for pure semidilute linear polymer solutions \cite{Hsiao2017}. The flow-strength dependence of ring polymer conformational fluctuations in ring-linear blends is further discussed in the section below. \\
	
	\textbf{D. Characteristic timescales of ring polymer conformational fluctuations}
	
	To further understand ring dynamics in ring-linear blends, we determined the autocorrelation of ring polymer extension fluctuations after the initial transient stretching phase. In particular, we used an autocorrelation analysis to quantify ring polymer extension fluctuations relative to the average polymer extension as a function of $Wi$. The autocorrelation function of a real-valued, integrable fluctuating quantity $x(t)$ is defined as:
	\begin{equation}
	C_{x,x}(\lambda)=\langle x(t)x(t+ \lambda ) \rangle_{t}
	\end{equation}
	where $\lambda$ is an offset time and $\langle \cdot \rangle_{t}$ denotes a time-averaged quantity. Here, fluctuations in ring polymer extensions are defined as the average (mean) extension $\langle l \rangle_t$ subtracted from the instantaneous chain extension $l(t)$. Thus, $l'(t)=l(t)-\langle l \rangle_t$ and the normalized autocorrelation function $C_{l^{'},l^{'}}$ is given by:
	\begin{equation}
	C_{l',l'}(\lambda) \equiv \frac{\langle l'(t) l'(t+\lambda) \rangle_t}{\langle l^{'2}(t) \rangle_t} = \frac{ \int_{-\infty}^{\infty} l'(t) l'(t+\lambda) dt}{\int_{-\infty}^{\infty} l^{'2}(t) dt}
	\end{equation}
	The quantity $C_{l',l'}$ is normalized by the autocorrelation function at zero offset time $\lambda=0$. The initial transient stretching phase (start-up phase) is discarded when calculating the autocorrelation function, similar to the calculation of $\langle \delta \rangle$ where polymer fractional extension is only considered between $t_{90}$ and $t_\mathrm{f}$, as described in Section III B. The offset time $\lambda$ is normalized by the strain rate $\dot{\epsilon}$.
	
		 \begin{figure*}
 			\includegraphics[scale=1.2]{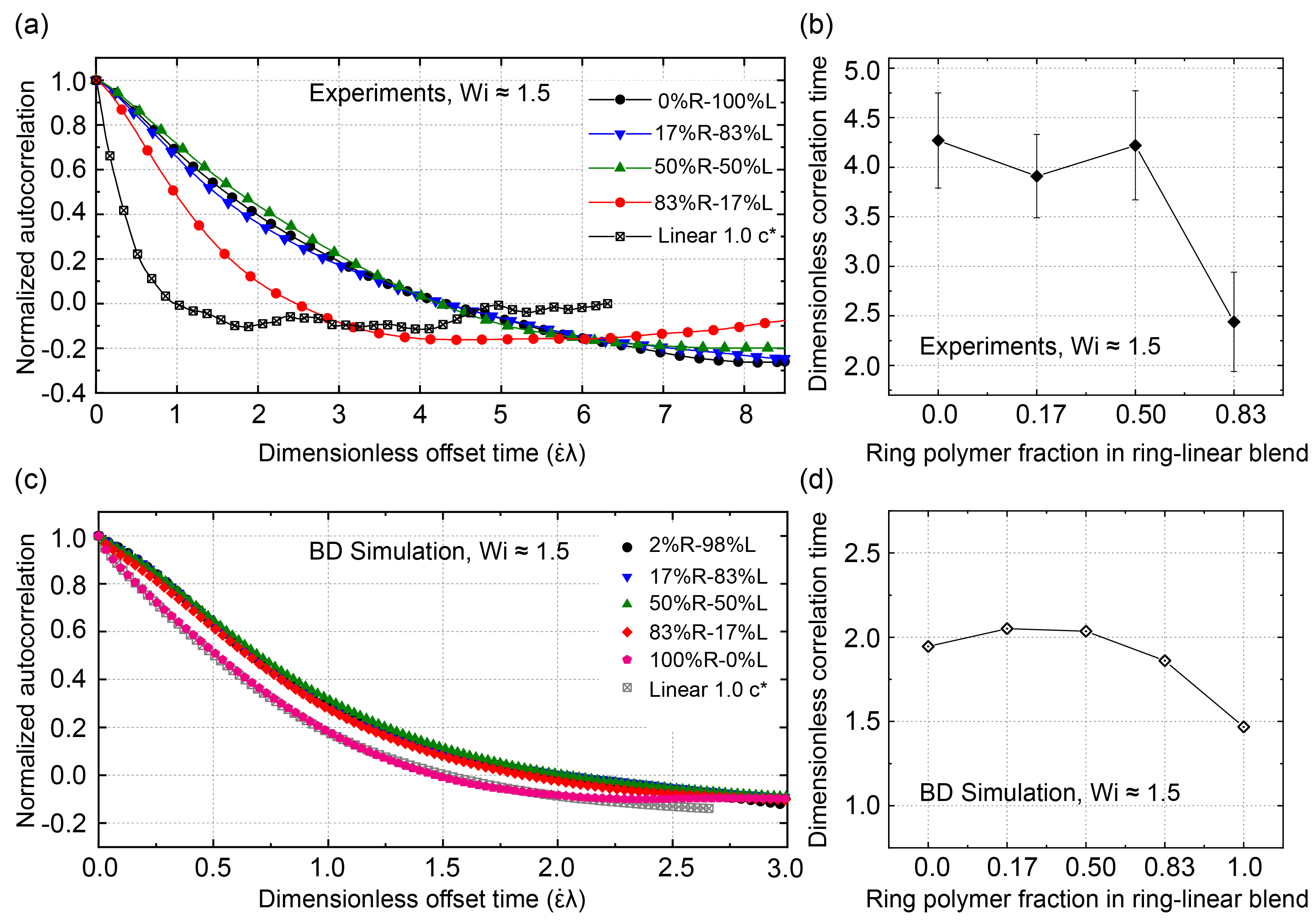}
  			\caption{Quantitative analysis of ring conformational fluctuations in flow. Autocorrelation of conformational fluctuations for ring polymers in semidilute ring-linear polymer blends after the initial start-up phase at $Wi \approx 1.5$ from (a) experiments and (c) BD simulations. Each experimental molecular ensemble contains $n \geq 34$ single molecule traces. The experimental autocorrelation of conformational fluctuations for pure linear semidilute polymer solutions (1 $c^*$, 50 $\mu$g/mL) is shown as a reference based on data from \citet{Hsiao2017}. Characteristic correlation times of ring polymer conformational fluctuations as a function of blend composition at $Wi \approx 1.5$ from (b) experiments and (d) BD simulations.}
  			\label{autocorr}
		\end{figure*}
	Fig. \ref{autocorr}a shows the autocorrelation functions of conformational fluctuations $C_{l',l'}$ for rings as a function of ring fraction in ring-linear blends at $Wi \approx 1.5$ from experiments. Upon increasing the ring fraction in blends from 0$\%$ R-100$\%$ L to 50$\%$ R-50$\%$ L, the autocorrelation function is relatively constant. However, the autocorrelation function rapidly decays when the fraction of rings increases in the 83$\%$ R-17$\%$ L blend. On the other hand, the autocorrelation function for linear polymers in pure linear semidilute solutions shows a more rapid decay compared to all of the ring-linear blends. In the case of pure linear semidilute solutions, fluctuations in fractional extension mainly arise due to Brownian fluctuations rather than intermolecular hooking interactions. The characteristic autocorrelation decay times are shown in Fig. \ref{autocorr}b, defined as the time at which the autocorrelation function in Fig. \ref{autocorr}a first equals zero. The dimensionless characteristic decay time (or decorrelation time) is approximately 4.2 strain units for blends with ring fraction between 0$\%$ R-100$\%$ L and 50$\%$ R-50$\%$ L, but significantly decreases to approximately 2.5 strain units for the 83$\%$ R-17$\%$ L blend.

    Autocorrelation functions of conformational fluctuations from molecular simulations are shown in Fig. \ref{autocorr}c, and  the characteristic autocorrelation decay times are determined as indicated in Fig. \ref{autocorr}d. Qualitative agreement is observed between simulations and SMFM experiments. The autocorrelation function shows similar behavior in blends from 2$\%$ R-98$\%$ L to 50$\%$ R-50$\%$ L, and begins to decay when the ring fraction increases to 83$\%$ R-17$\%$ L. The quickest decay in autocorrelation functions occurs in pure semidilute ring (100$\%$ R-0$\%$ L) and linear polymer solutions and is more rapid than any ring-linear polymer blend solutions.	
    
	Our results indicate that, in an average sense, the interaction timescale between ring and linear polymer chains remains relatively constant when the majority of polymers in ring-linear blends is linear. However, it is important to note that due to the role of stochasticity, one would expect a distribution of molecular sub-populations over the entire ensemble. From this view, the ensemble-averaged autocorrelation functions may not reflect the drastic variations in molecular extension that may occur in individual molecules, as shown in Fig. \ref{SMsnapshots}. Rather, the autocorrelation functions reflect the average intermolecular interaction timescales between rings and adjacent ring or linear polymers in blends. \\
	
	\textbf{E. Average fractional extension of rings}
		 \begin{figure}
 			\includegraphics[scale=1.0]{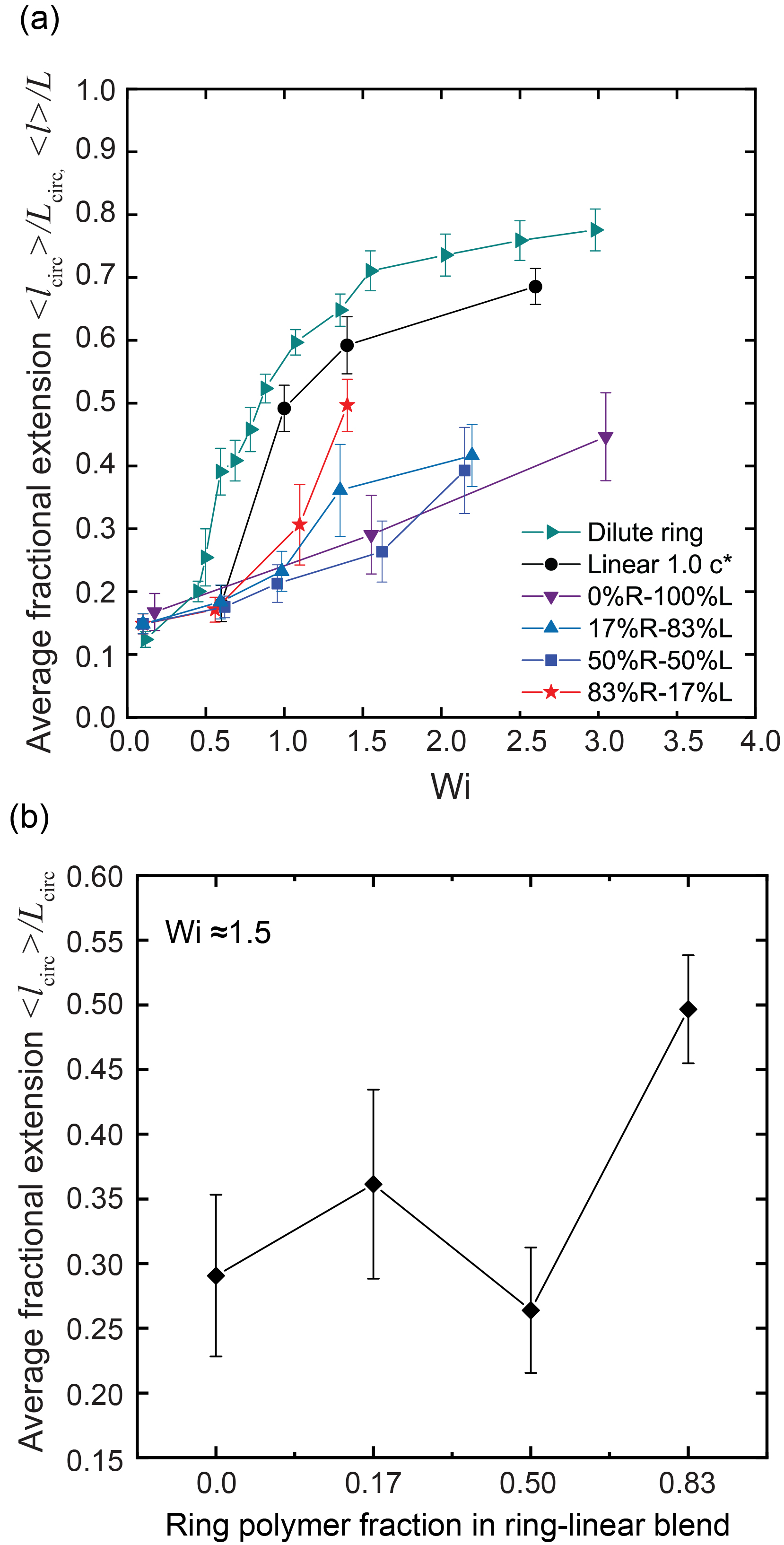}
  			\caption{Steady fractional extension in extensional flow. (a) Steady fractional extension of pure semidilute linear polymer solutions at 1 c$^*$ (50 $\mu$g/mL) and ring polymers in 50 $\mu$g/mL semidilute ring-linear blends in extensional flow. Experimental data for dilute ring polymers are taken from \citet{Li2015}, and data for 1 c$^*$ linear polymers are taken from \citet{Hsiao2017}. (b) Average steady fraction extension of ring polymers in different semidilute ring-linear polymer blend at $Wi \approx 1.5$. Each molecular ensemble consists of at least $n \geq$ 34 single molecule trajectories.}
  			\label{avg_extension}
		\end{figure}
	
    We further determined the average fractional extension for ring polymers in semidilute ring-linear polymer blends, using a method similar to determining the average fractional extension of polymers under large amplitude oscillatory extensional (LAOE) flow \cite{Zhou2016,Zhou2016b}. In brief, we define an average steady fractional extension between $t_{90}$ and $t_f$, determined from the fractional extension after the initial transient start-up phase, where $t_\mathrm{f}$ is the time at which the step strain rate input stops and $t_{90}$ is the time at which the fractional polymer extension first reaches 90$\%$ of the average fractional extension at time $t_\mathrm{f}$. In general, this method corresponds to determining average polymer extension for fluid strains approximately $\epsilon > 8$ in steady extensional flow. The average fractional extension for ring polymers in dilute solution \cite{Li2015} and for linear polymers in semidilute solution \cite{Hsiao2017} are plotted as a reference (Fig. \ref{avg_extension}a). Prior work has shown that ring polymers exhibit a delayed coil-stretch transition in dilute solutions due to intramolecular HI between the two strands in extensional flow \cite{Li2015, Hsiao2016a}, which is not observed for linear polymers in dilute solutions \cite{Perkins1997}. Moreover, linear polymers show a slight increase in the critical Weissenberg number at the coil-stretch transition, $Wi_c$, in semidilute unentangled solutions due to intermolecular HI and additional molecular interactions between the neighboring chains.
	
	Interestingly, our results show that $Wi_c$ for ring polymers depends on the composition of semidilute ring-linear blends, even for the same total polymer concentration of 50 $\mu$g/mL. We estimate $Wi_c$ based on the average fractional extension for ring polymers in semidilute ring-linear polymer blends at the coil-stretch transition, $\langle \tilde{l}\mathrm{_c} \rangle$. Here, $\langle \tilde{l}\mathrm{_c} \rangle = \langle l\mathrm{_c} \rangle / L_\mathrm{{circ}}$ is determined as the mean value between the coiled and stretched limits in a logarithmic scale, as previously described \cite{HernandezCifre2001}. Hence, $\langle \tilde{l}\mathrm{_c} \rangle$ is defined as
	\begin{equation}
	ln{\langle \tilde{l}\mathrm{_c} \rangle}^2 = (ln{\langle \tilde{l}_0 \rangle}^2 + ln{\langle \tilde{l}_\mathrm{max} \rangle}^2)/2
	\end{equation}
	where $\tilde{l}_0 = \langle l_0 \rangle / L_\mathrm{{circ}}$ is the fractional extension of ring polymers in the equilibrium coiled state and $\tilde{l}_\mathrm{max} = \langle l_\mathrm{max} \rangle / L_\mathrm{{circ}}$ is the maximum fractional extension for ring polymers in our experiments. In this way, the critical Weissenberg number at the coil-stretch transition is determined by finding the corresponding Weissenberg number at $\langle \tilde{l}\mathrm{_c} \rangle$, such that $Wi\mathrm{_c}$ = 1.5, $Wi\mathrm{_c}$ = 1.1, $Wi\mathrm{_c}$ = 1.4, and $Wi\mathrm{_c}$ = 0.9 for ring polymers in 0$\%$ R-100$\%$ L, 17$\%$ R-83$\%$ L, 50$\%$ R-50$\%$ L, and 83$\%$ R-17$\%$ L  ring-linear polymer blends, respectively. The $Wi\mathrm{_c}$ values are in reasonable agreement with the Weissenberg numbers corresponding to the maximum fluctuation quantities $\langle \delta \rangle$  (Fig. \ref{conf_fluc}), as discussed in Section III C. 
	
		 \begin{figure}
 			\includegraphics[scale=1]{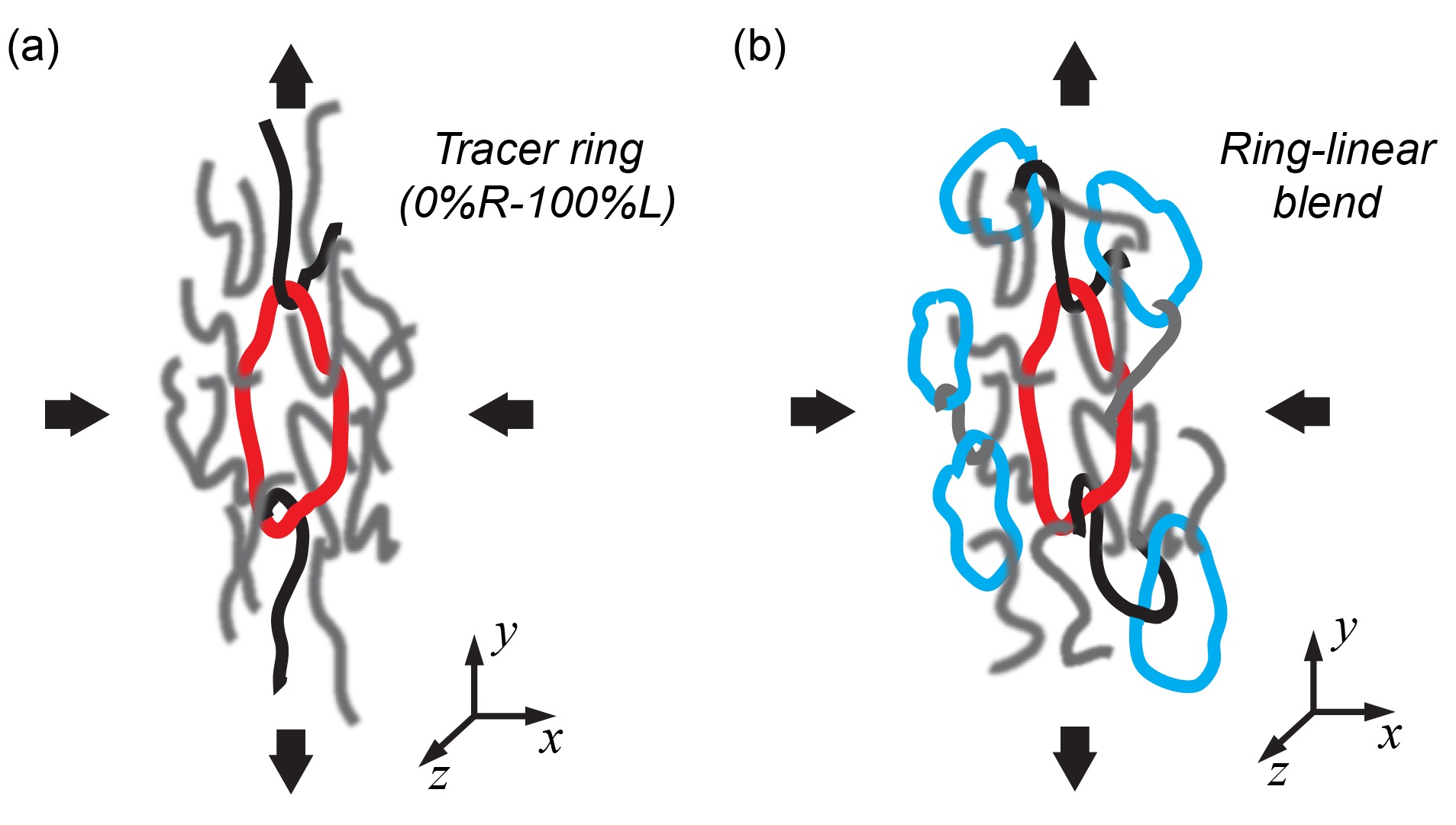}
  			\caption{Schematic of intermolecular interactions in semidilute ring-linear blends. (a) In nearly pure semidilute linear polymer solutions (0$\%$ R-100$\%$ L), background linear polymers thread into open ring polymers in extensional flow. Threaded states are not necessarily limited to a doubly-threaded state as shown here. (b) In semidilute ring-linear polymer blends, linear polymers can thread with multiple rings, potentially leading to a local transient interlinked structure in flow.}
  			\label{Diff_scheme}
		\end{figure}
	
	Based on these results, ring polymer dynamics are influenced by both polymer architecture and the relative composition in the blends, thereby affecting both the critical Weissenberg number at the coil-stretch transition and the average fractional extension. Interestingly, the 50$\%$ R-50$\%$ L blend shares the largest $Wi\mathrm{_c}$ with pure linear polymer solutions (0$\%$ R-100$\%$ L) but shows the smallest average fractional extension. Fig. \ref{avg_extension}b further shows a direct comparison of average fractional extension as a function of ring-linear blend composition at $Wi \approx 1.5$. Broadly, these results are consistent with recent microrheology experiments on concentrated ring-linear blends of DNA, where the largest plateau modulus was observed for blends containing comparable amounts of ring and linear polymers \cite{Peddireddy2020a}. Taken together, our results support a scenario in which ring and linear polymers strongly interact in semidilute ring-linear blend solutions (Fig. \ref{Diff_scheme}), with clear differences compared to ultra-dilute solutions or pure linear semidilute solutions.  

\section{Conclusions}

	Understanding ring polymer dynamics is a particularly challenging and interesting problem in soft materials and rheology. Despite recent progress, we do not yet fully understand the combined roles of molecular architecture, intermolecular interactions, and long-range HI on the dynamics of ring-linear blends under \textit{nonequilibrium} conditions. In this work, we use single molecule fluorescence microscopy coupled with automated flow control and microfluidics to systematically investigate the nonequilibrium dynamics of ring polymers in semidilute ring-linear DNA blend solutions. Our results show molecular evidence of large conformational fluctuations of ring polymers in steady flows, which arise due to a combination of linear chain threading into open rings and strong intermolecular HI in flow \cite{young2020dynamics}. Our results are consistent with a molecular picture wherein strongly-interacting ring-linear transient structures form and may exhibit local resistance to stretching in flow, especially when the blend contains comparable amounts of ring and linear polymers.  

	The relaxation dynamics of ring DNA polymers in semidilute ring-linear blends is governed by two distinct molecular sub-populations. One sub-population exhibits single-mode exponential relaxation behavior, which is attributed to the relaxation of isolated ring polymers that are not associated with intermolecular transient structures between ring and linear polymers following the cessation of flow. The emergence of a second molecular sub-population showing a double-mode relaxation response likely arises due to interactions of ring polymers with background linear polymers, including ring-linear chain threading and solvent-mediated HI effects. The probability of double-mode exponential relaxation first increases with increasing the ring fraction in the blend, followed by a decrease with increasing ring fraction up to the 83$\%$ R-17$\%$ L blend. These results indicate that different blend compositions alter the degree of interchain interactions between ring and linear polymers, thereby affecting the relaxation dynamics of rings.
	
	Our results show strikingly large conformational fluctuations for rings in ring-linear blends in steady extensional flow. By quantifying the conformational fluctuations through a chain fluctuation quantity $\langle \delta \rangle$, our results show that chain fluctuations increase with ring polymer fraction in blends but substantially decrease when the blend contains $>80\%$ of rings. In addition, our results reveal a unique set of molecular conformations and a marked increase in molecular individualism during the transient stretching process for rings in 17$\%$ R-83$\%$ L and 50$\%$ R-50$\%$ L blends. Surprisingly, individual rings are observed to tumble, re-coil, and re-stretch in semidilute ring-linear blends in planar extensional flow, a behavior only observed previously for dilute linear \cite{Smith1999, Schroeder2018} and ring polymers \cite{Tu2020} under simple shear flow. These behaviors are attributed to a combination of long-range intermolecular HI in semi-dilute solutions, in addition to chain-chain interactions in flow. Simulations results directly capture these conformational fluctuations and provide further evidence that HI plays an important role in semidilute solutions \cite{young2020dynamics}. The autocorrelation of ring polymer fluctuations shows a slower rate of decay and a longer correlation time for ring polymers in 0$\%$ R-100$\%$ L, 17$\%$ R-83$\%$ L and 50$\%$ R-50$\%$ L blends. Hence, we hypothesize that large conformational fluctuations and the unique molecular individualism are indicative of linear polymers threading into rings to form transient intermolecular structures in flow (Fig. \ref{Diff_scheme}). 
	
	Our results further show a dependence on molecular stretching and conformation as a function of the ring fraction in ring-linear blends. As the fraction of ring polymers increases in the 83$\%$ R-17$\%$ L blend, ring chain extension fluctuations sharply decrease, nearly resembling fluctuations of linear polymers in pure semidilute linear solutions \cite{Hsiao2017}. The small magnitude conformational fluctuations in fractional extension also result in a large average fractional extension, and the autocorrelation function of the conformational fluctuation decays noticeably faster for ring polymers in 83$\%$ R-17$\%$ L than the other three blends. Coarse-grained molecular simulations show that reduced ring fluctuations in semidilute ring-linear blends with high ring fraction occurs due to decreased intermolecular HI effects between highly stretched linear chains and adjacent ring polymers \cite{young2020dynamics}.
	
	Taken together, these results provide an new molecular understanding of ring polymer dynamics in ring-linear blends. In particular, our combined experimental and computational results show direct molecular evidence for the transient threading of linear polymers through open ring polymers in flow. From a broad perspective, this work provides an improved understanding of ring dynamics in non-dilute polymer solutions, revealing new information regarding the nonequilibrium dynamics of rings that may be useful informing the future design and processing of polymer solutions with complex molecular architectures.
	
\begin{acknowledgments}
This research was supported by the National Science Foundation (NSF) Award CBET-1604038 (Y.Z. and C.M.S.) and partially supported by the NSF through the University of Illinois at Urbana-Champaign Materials Research Science and Engineering Center (MRSEC) DMR-1720633 (Y.Z. and C.M.S.), a PPG-MRL graduate research assistantship award (Y.Z.), a DuPont Science \& Engineering fellowship (C.D.Y), NSF Award CBET-1803757 (C.D.Y. and C.E.S.) NSF Award CBET-1603925 (K.E.R., M.L., and R.M.R-A.), and NSF Award CBET-1603943 (S.B., D.K, and G.B.M.).
\end{acknowledgments}



\begin{thebibliography}{72}%
\makeatletter
\providecommand \@ifxundefined [1]{%
 \@ifx{#1\undefined}
}%
\providecommand \@ifnum [1]{%
 \ifnum #1\expandafter \@firstoftwo
 \else \expandafter \@secondoftwo
 \fi
}%
\providecommand \@ifx [1]{%
 \ifx #1\expandafter \@firstoftwo
 \else \expandafter \@secondoftwo
 \fi
}%
\providecommand \natexlab [1]{#1}%
\providecommand \enquote  [1]{``#1''}%
\providecommand \bibnamefont  [1]{#1}%
\providecommand \bibfnamefont [1]{#1}%
\providecommand \citenamefont [1]{#1}%
\providecommand \href@noop [0]{\@secondoftwo}%
\providecommand \href [0]{\begingroup \@sanitize@url \@href}%
\providecommand \@href[1]{\@@startlink{#1}\@@href}%
\providecommand \@@href[1]{\endgroup#1\@@endlink}%
\providecommand \@sanitize@url [0]{\catcode `\\12\catcode `\$12\catcode
  `\&12\catcode `\#12\catcode `\^12\catcode `\_12\catcode `\%12\relax}%
\providecommand \@@startlink[1]{}%
\providecommand \@@endlink[0]{}%
\providecommand \url  [0]{\begingroup\@sanitize@url \@url }%
\providecommand \@url [1]{\endgroup\@href {#1}{\urlprefix }}%
\providecommand \urlprefix  [0]{URL }%
\providecommand \Eprint [0]{\href }%
\providecommand \doibase [0]{https://doi.org/}%
\providecommand \selectlanguage [0]{\@gobble}%
\providecommand \bibinfo  [0]{\@secondoftwo}%
\providecommand \bibfield  [0]{\@secondoftwo}%
\providecommand \translation [1]{[#1]}%
\providecommand \BibitemOpen [0]{}%
\providecommand \bibitemStop [0]{}%
\providecommand \bibitemNoStop [0]{.\EOS\space}%
\providecommand \EOS [0]{\spacefactor3000\relax}%
\providecommand \BibitemShut  [1]{\csname bibitem#1\endcsname}%
\let\auto@bib@innerbib\@empty
\bibitem [{\citenamefont {McLeish}(2002)}]{McLeish2002}%
  \BibitemOpen
  \bibfield  {author} {\bibinfo {author} {\bibfnamefont {T.~C.~B.}\
  \bibnamefont {McLeish}},\ }\bibfield  {title} {\bibinfo {title} {{Polymers
  without beginning or end}},\ }\href {https://doi.org/10.1126/science.1076810}
  {\bibfield  {journal} {\bibinfo  {journal} {Science}\ }\textbf {\bibinfo
  {volume} {297}},\ \bibinfo {pages} {2005} (\bibinfo {year}
  {2002})}\BibitemShut {NoStop}%
\bibitem [{\citenamefont {Taanman}(1999)}]{Taanman1999}%
  \BibitemOpen
  \bibfield  {author} {\bibinfo {author} {\bibfnamefont {J.~W.}\ \bibnamefont
  {Taanman}},\ }\bibfield  {title} {\bibinfo {title} {{The mitochondrial
  genome: Structure, transcription, translation and replication}},\ }\href
  {https://doi.org/10.1016/S0005-2728(98)00161-3} {\bibfield  {journal}
  {\bibinfo  {journal} {Biochimica et Biophysica Acta - Bioenergetics}\
  }\textbf {\bibinfo {volume} {1410}},\ \bibinfo {pages} {103} (\bibinfo {year}
  {1999})}\BibitemShut {NoStop}%
\bibitem [{\citenamefont {Halverson}\ \emph {et~al.}(2014)\citenamefont
  {Halverson}, \citenamefont {Smrek}, \citenamefont {Kremer},\ and\
  \citenamefont {Grosberg}}]{Halverson2014}%
  \BibitemOpen
  \bibfield  {author} {\bibinfo {author} {\bibfnamefont {J.~D.}\ \bibnamefont
  {Halverson}}, \bibinfo {author} {\bibfnamefont {J.}~\bibnamefont {Smrek}},
  \bibinfo {author} {\bibfnamefont {K.}~\bibnamefont {Kremer}},\ and\ \bibinfo
  {author} {\bibfnamefont {A.~Y.}\ \bibnamefont {Grosberg}},\ }\bibfield
  {title} {\bibinfo {title} {{From a melt of rings to chromosome territories:
  The role of topological constraints in genome folding}},\ }\href
  {https://doi.org/10.1088/0034-4885/77/2/022601} {\bibfield  {journal}
  {\bibinfo  {journal} {Reports on Progress in Physics}\ }\textbf {\bibinfo
  {volume} {77}},\ \bibinfo {pages} {022601} (\bibinfo {year} {2014})},\
  \Eprint {https://arxiv.org/abs/1311.5262} {arXiv:1311.5262} \BibitemShut
  {NoStop}%
\bibitem [{\citenamefont {Deutman}\ \emph {et~al.}(2008)\citenamefont
  {Deutman}, \citenamefont {Monnereau}, \citenamefont {Elemans}, \citenamefont
  {Ercolani}, \citenamefont {Nolte},\ and\ \citenamefont
  {Rowan}}]{Deutman2008}%
  \BibitemOpen
  \bibfield  {author} {\bibinfo {author} {\bibfnamefont {A.~B.~C.}\
  \bibnamefont {Deutman}}, \bibinfo {author} {\bibfnamefont {C.}~\bibnamefont
  {Monnereau}}, \bibinfo {author} {\bibfnamefont {J.~A. A.~W.}\ \bibnamefont
  {Elemans}}, \bibinfo {author} {\bibfnamefont {G.}~\bibnamefont {Ercolani}},
  \bibinfo {author} {\bibfnamefont {R.~J.~M.}\ \bibnamefont {Nolte}},\ and\
  \bibinfo {author} {\bibfnamefont {A.~E.}\ \bibnamefont {Rowan}},\ }\bibfield
  {title} {\bibinfo {title} {{Mechanism of threading a polymer through a
  macrocyclic ring}},\ }\href {https://doi.org/10.1126/science.1164647}
  {\bibfield  {journal} {\bibinfo  {journal} {Science}\ }\textbf {\bibinfo
  {volume} {322}},\ \bibinfo {pages} {1668} (\bibinfo {year}
  {2008})}\BibitemShut {NoStop}%
\bibitem [{\citenamefont {Edwards}\ \emph {et~al.}(2019)\citenamefont
  {Edwards}, \citenamefont {Wolf},\ and\ \citenamefont {Grubbs}}]{Edwards2019}%
  \BibitemOpen
  \bibfield  {author} {\bibinfo {author} {\bibfnamefont {J.~P.}\ \bibnamefont
  {Edwards}}, \bibinfo {author} {\bibfnamefont {W.~J.}\ \bibnamefont {Wolf}},\
  and\ \bibinfo {author} {\bibfnamefont {R.~H.}\ \bibnamefont {Grubbs}},\
  }\bibfield  {title} {\bibinfo {title} {{The synthesis of cyclic polymers by
  olefin metathesis: Achievements and challenges}},\ }\href
  {https://doi.org/10.1002/pola.29253} {\bibfield  {journal} {\bibinfo
  {journal} {Journal of Polymer Science, Part A: Polymer Chemistry}\ }\textbf
  {\bibinfo {volume} {57}},\ \bibinfo {pages} {228} (\bibinfo {year}
  {2019})}\BibitemShut {NoStop}%
\bibitem [{\citenamefont {Feinberg}\ \emph {et~al.}(2018)\citenamefont
  {Feinberg}, \citenamefont {Hernandez}, \citenamefont {Plantz}, \citenamefont
  {Mejia}, \citenamefont {Sottos}, \citenamefont {White},\ and\ \citenamefont
  {Moore}}]{Feinberg2018}%
  \BibitemOpen
  \bibfield  {author} {\bibinfo {author} {\bibfnamefont {A.~M.}\ \bibnamefont
  {Feinberg}}, \bibinfo {author} {\bibfnamefont {H.~L.}\ \bibnamefont
  {Hernandez}}, \bibinfo {author} {\bibfnamefont {C.~L.}\ \bibnamefont
  {Plantz}}, \bibinfo {author} {\bibfnamefont {E.~B.}\ \bibnamefont {Mejia}},
  \bibinfo {author} {\bibfnamefont {N.~R.}\ \bibnamefont {Sottos}}, \bibinfo
  {author} {\bibfnamefont {S.~R.}\ \bibnamefont {White}},\ and\ \bibinfo
  {author} {\bibfnamefont {J.~S.}\ \bibnamefont {Moore}},\ }\bibfield  {title}
  {\bibinfo {title} {{Cyclic poly(phthalaldehyde): thermoforming a bulk
  transient material}},\ }\href {https://doi.org/10.1021/acsmacrolett.7b00769}
  {\bibfield  {journal} {\bibinfo  {journal} {ACS Macro Letters}\ }\textbf
  {\bibinfo {volume} {7}},\ \bibinfo {pages} {47} (\bibinfo {year}
  {2018})}\BibitemShut {NoStop}%
\bibitem [{\citenamefont {Lloyd}\ \emph {et~al.}(2019)\citenamefont {Lloyd},
  \citenamefont {Hernandez}, \citenamefont {Feinberg}, \citenamefont
  {Yourdkhani}, \citenamefont {Zen}, \citenamefont {Mejia}, \citenamefont
  {Sottos}, \citenamefont {Moore},\ and\ \citenamefont {White}}]{Lloyd2019}%
  \BibitemOpen
  \bibfield  {author} {\bibinfo {author} {\bibfnamefont {E.~M.}\ \bibnamefont
  {Lloyd}}, \bibinfo {author} {\bibfnamefont {H.~L.}\ \bibnamefont
  {Hernandez}}, \bibinfo {author} {\bibfnamefont {A.~M.}\ \bibnamefont
  {Feinberg}}, \bibinfo {author} {\bibfnamefont {M.}~\bibnamefont
  {Yourdkhani}}, \bibinfo {author} {\bibfnamefont {E.~K.}\ \bibnamefont {Zen}},
  \bibinfo {author} {\bibfnamefont {E.~B.}\ \bibnamefont {Mejia}}, \bibinfo
  {author} {\bibfnamefont {N.~R.}\ \bibnamefont {Sottos}}, \bibinfo {author}
  {\bibfnamefont {J.~S.}\ \bibnamefont {Moore}},\ and\ \bibinfo {author}
  {\bibfnamefont {S.~W.}\ \bibnamefont {White}},\ }\bibfield  {title} {\bibinfo
  {title} {Fully recyclable metastable polymers and composites},\ }\href@noop
  {} {\bibfield  {journal} {\bibinfo  {journal} {Chemistry of Materials}\
  }\textbf {\bibinfo {volume} {31}},\ \bibinfo {pages} {398} (\bibinfo {year}
  {2019})}\BibitemShut {NoStop}%
\bibitem [{\citenamefont {Rosenthal-Kim}\ and\ \citenamefont
  {Puskas}(2015)}]{Puskas2015}%
  \BibitemOpen
  \bibfield  {author} {\bibinfo {author} {\bibfnamefont {E.~Q.}\ \bibnamefont
  {Rosenthal-Kim}}\ and\ \bibinfo {author} {\bibfnamefont {J.~E.}\ \bibnamefont
  {Puskas}},\ }\bibfield  {title} {\bibinfo {title} {Green polymer chemistry:
  investigating the mechanism of radical ring-opening redox polymerization
  ({R3P}) of 3,6-dioxa-1,8-octanedithiol ({DODT})},\ }\href@noop {} {\bibfield
  {journal} {\bibinfo  {journal} {Molecules}\ }\textbf {\bibinfo {volume}
  {20}},\ \bibinfo {pages} {6504–6519} (\bibinfo {year} {2015})}\BibitemShut
  {NoStop}%
\bibitem [{\citenamefont {Rosenthal-Kim}\ and\ \citenamefont
  {Puskas}(2012)}]{Puskas2012}%
  \BibitemOpen
  \bibfield  {author} {\bibinfo {author} {\bibfnamefont {E.}~\bibnamefont
  {Rosenthal-Kim}}\ and\ \bibinfo {author} {\bibfnamefont {J.~E.}\ \bibnamefont
  {Puskas}},\ }\bibfield  {title} {\bibinfo {title} {Green polymer chemistry:
  living oxidative polymerization of dithiols},\ }\href@noop {} {\bibfield
  {journal} {\bibinfo  {journal} {Pure Appl. Chem.}\ }\textbf {\bibinfo
  {volume} {84}},\ \bibinfo {pages} {2121–2133} (\bibinfo {year}
  {2012})}\BibitemShut {NoStop}%
\bibitem [{\citenamefont {Roovers}(1985)}]{Roovers1985}%
  \BibitemOpen
  \bibfield  {author} {\bibinfo {author} {\bibfnamefont {J.}~\bibnamefont
  {Roovers}},\ }\bibfield  {title} {\bibinfo {title} {{Melt Properties of Ring
  Polystyrenes}},\ }\href {https://doi.org/10.1021/ma00148a059} {\bibfield
  {journal} {\bibinfo  {journal} {Macomolecules}\ }\textbf {\bibinfo {volume}
  {18}},\ \bibinfo {pages} {1359} (\bibinfo {year} {1985})}\BibitemShut
  {NoStop}%
\bibitem [{\citenamefont {Mckenna}\ \emph {et~al.}(1987)\citenamefont
  {Mckenna}, \citenamefont {Hadziioannou}, \citenamefont {Lutz}, \citenamefont
  {Hild}, \citenamefont {Strazielle}, \citenamefont {Straupe}, \citenamefont
  {Rempp},\ and\ \citenamefont {Kovacs}}]{McKenna1987}%
  \BibitemOpen
  \bibfield  {author} {\bibinfo {author} {\bibfnamefont {G.~B.}\ \bibnamefont
  {Mckenna}}, \bibinfo {author} {\bibfnamefont {G.}~\bibnamefont
  {Hadziioannou}}, \bibinfo {author} {\bibfnamefont {P.}~\bibnamefont {Lutz}},
  \bibinfo {author} {\bibfnamefont {G.}~\bibnamefont {Hild}}, \bibinfo {author}
  {\bibfnamefont {C.}~\bibnamefont {Strazielle}}, \bibinfo {author}
  {\bibfnamefont {C.}~\bibnamefont {Straupe}}, \bibinfo {author} {\bibfnamefont
  {P.}~\bibnamefont {Rempp}},\ and\ \bibinfo {author} {\bibfnamefont {A.~J.}\
  \bibnamefont {Kovacs}},\ }\bibfield  {title} {\bibinfo {title} {{Dilute
  Solution Characterization of Cyclic Polystyrene Molecules and Their
  Zero-Shear Viscosity in the Melt}},\ }\href
  {https://doi.org/10.1021/ma00169a007} {\bibfield  {journal} {\bibinfo
  {journal} {Macromelecules}\ }\textbf {\bibinfo {volume} {20}},\ \bibinfo
  {pages} {498} (\bibinfo {year} {1987})}\BibitemShut {NoStop}%
\bibitem [{\citenamefont {Roovers}(1988)}]{Roovers1988}%
  \BibitemOpen
  \bibfield  {author} {\bibinfo {author} {\bibfnamefont {J.}~\bibnamefont
  {Roovers}},\ }\bibfield  {title} {\bibinfo {title} {{Viscoelastic properties
  of polybutadiene rings}},\ }\href {https://doi.org/10.1021/ma00183a049}
  {\bibfield  {journal} {\bibinfo  {journal} {Macromolecules}\ }\textbf
  {\bibinfo {volume} {21}},\ \bibinfo {pages} {1517} (\bibinfo {year}
  {1988})}\BibitemShut {NoStop}%
\bibitem [{\citenamefont {McKenna}\ \emph {et~al.}(1989)\citenamefont
  {McKenna}, \citenamefont {Hostetter}, \citenamefont {Hadjichristidis},
  \citenamefont {Fetters},\ and\ \citenamefont {Plazek}}]{Mckenna1989}%
  \BibitemOpen
  \bibfield  {author} {\bibinfo {author} {\bibfnamefont {G.~B.}\ \bibnamefont
  {McKenna}}, \bibinfo {author} {\bibfnamefont {B.~J.}\ \bibnamefont
  {Hostetter}}, \bibinfo {author} {\bibfnamefont {N.}~\bibnamefont
  {Hadjichristidis}}, \bibinfo {author} {\bibfnamefont {L.~J.}\ \bibnamefont
  {Fetters}},\ and\ \bibinfo {author} {\bibfnamefont {D.~J.}\ \bibnamefont
  {Plazek}},\ }\bibfield  {title} {\bibinfo {title} {{A study of the linear
  viscoelastic properties of cyclic polystyrenes using creep and recovery
  measurements}},\ }\href {https://doi.org/10.1021/ma00194a056} {\bibfield
  {journal} {\bibinfo  {journal} {Macromelecules}\ }\textbf {\bibinfo {volume}
  {22}},\ \bibinfo {pages} {1834} (\bibinfo {year} {1989})}\BibitemShut
  {NoStop}%
\bibitem [{\citenamefont {Halverson}\ \emph {et~al.}(2011)\citenamefont
  {Halverson}, \citenamefont {Lee}, \citenamefont {Grest}, \citenamefont
  {Grosberg},\ and\ \citenamefont {Kremer}}]{Halverson2011b}%
  \BibitemOpen
  \bibfield  {author} {\bibinfo {author} {\bibfnamefont {J.~D.}\ \bibnamefont
  {Halverson}}, \bibinfo {author} {\bibfnamefont {W.~B.}\ \bibnamefont {Lee}},
  \bibinfo {author} {\bibfnamefont {G.~S.}\ \bibnamefont {Grest}}, \bibinfo
  {author} {\bibfnamefont {A.~Y.}\ \bibnamefont {Grosberg}},\ and\ \bibinfo
  {author} {\bibfnamefont {K.}~\bibnamefont {Kremer}},\ }\bibfield  {title}
  {\bibinfo {title} {{Molecular dynamics simulation study of nonconcatenated
  ring polymers in a melt. I. Statics}},\ }\href
  {https://doi.org/10.1063/1.3587137} {\bibfield  {journal} {\bibinfo
  {journal} {Journal of Chemical Physics}\ }\textbf {\bibinfo {volume} {134}},\
  \bibinfo {pages} {204904} (\bibinfo {year} {2011})},\ \Eprint
  {https://arxiv.org/abs/arXiv:1104.5653v1} {arXiv:arXiv:1104.5653v1}
  \BibitemShut {NoStop}%
\bibitem [{\citenamefont {Pasquino}\ \emph {et~al.}(2013)\citenamefont
  {Pasquino}, \citenamefont {Vasilakopoulos}, \citenamefont {Jeong},
  \citenamefont {Lee}, \citenamefont {Rogers}, \citenamefont {Sakellariou},
  \citenamefont {Allgaier}, \citenamefont {Takano}, \citenamefont {Br{\'{a}}s},
  \citenamefont {Chang}, \citenamefont {Goo{\ss}en}, \citenamefont
  {Pyckhout-Hintzen}, \citenamefont {Wischnewski}, \citenamefont
  {Hadjichristidis}, \citenamefont {Richter}, \citenamefont {Rubinstein},\ and\
  \citenamefont {Vlassopoulos}}]{Pasquino2013a}%
  \BibitemOpen
  \bibfield  {author} {\bibinfo {author} {\bibfnamefont {R.}~\bibnamefont
  {Pasquino}}, \bibinfo {author} {\bibfnamefont {T.~C.}\ \bibnamefont
  {Vasilakopoulos}}, \bibinfo {author} {\bibfnamefont {Y.~C.}\ \bibnamefont
  {Jeong}}, \bibinfo {author} {\bibfnamefont {H.}~\bibnamefont {Lee}}, \bibinfo
  {author} {\bibfnamefont {S.}~\bibnamefont {Rogers}}, \bibinfo {author}
  {\bibfnamefont {G.}~\bibnamefont {Sakellariou}}, \bibinfo {author}
  {\bibfnamefont {J.}~\bibnamefont {Allgaier}}, \bibinfo {author}
  {\bibfnamefont {A.}~\bibnamefont {Takano}}, \bibinfo {author} {\bibfnamefont
  {A.~R.}\ \bibnamefont {Br{\'{a}}s}}, \bibinfo {author} {\bibfnamefont
  {T.}~\bibnamefont {Chang}}, \bibinfo {author} {\bibfnamefont
  {S.}~\bibnamefont {Goo{\ss}en}}, \bibinfo {author} {\bibfnamefont
  {W.}~\bibnamefont {Pyckhout-Hintzen}}, \bibinfo {author} {\bibfnamefont
  {A.}~\bibnamefont {Wischnewski}}, \bibinfo {author} {\bibfnamefont
  {N.}~\bibnamefont {Hadjichristidis}}, \bibinfo {author} {\bibfnamefont
  {D.}~\bibnamefont {Richter}}, \bibinfo {author} {\bibfnamefont
  {M.}~\bibnamefont {Rubinstein}},\ and\ \bibinfo {author} {\bibfnamefont
  {D.}~\bibnamefont {Vlassopoulos}},\ }\bibfield  {title} {\bibinfo {title}
  {{Viscosity of ring polymer melts}},\ }\href
  {https://doi.org/10.1021/mz400344e} {\bibfield  {journal} {\bibinfo
  {journal} {ACS Macro Letters}\ }\textbf {\bibinfo {volume} {2}},\ \bibinfo
  {pages} {874} (\bibinfo {year} {2013})}\BibitemShut {NoStop}%
\bibitem [{\citenamefont {Kapnistos}\ \emph {et~al.}(2008)\citenamefont
  {Kapnistos}, \citenamefont {Lang}, \citenamefont {Vlassopoulos},
  \citenamefont {Pyckhout-Hintzen}, \citenamefont {Richter}, \citenamefont
  {Cho}, \citenamefont {Chang},\ and\ \citenamefont
  {Rubinstein}}]{Kapnistos2008a}%
  \BibitemOpen
  \bibfield  {author} {\bibinfo {author} {\bibfnamefont {M.}~\bibnamefont
  {Kapnistos}}, \bibinfo {author} {\bibfnamefont {M.}~\bibnamefont {Lang}},
  \bibinfo {author} {\bibfnamefont {D.}~\bibnamefont {Vlassopoulos}}, \bibinfo
  {author} {\bibfnamefont {W.}~\bibnamefont {Pyckhout-Hintzen}}, \bibinfo
  {author} {\bibfnamefont {D.}~\bibnamefont {Richter}}, \bibinfo {author}
  {\bibfnamefont {D.}~\bibnamefont {Cho}}, \bibinfo {author} {\bibfnamefont
  {T.}~\bibnamefont {Chang}},\ and\ \bibinfo {author} {\bibfnamefont
  {M.}~\bibnamefont {Rubinstein}},\ }\bibfield  {title} {\bibinfo {title}
  {{Unexpected power-law stress relaxation of entangled ring polymers.}},\
  }\href {https://doi.org/10.1038/nmat2292} {\bibfield  {journal} {\bibinfo
  {journal} {Nature Materials}\ }\textbf {\bibinfo {volume} {7}},\ \bibinfo
  {pages} {997} (\bibinfo {year} {2008})}\BibitemShut {NoStop}%
\bibitem [{\citenamefont {Doi}\ \emph {et~al.}(2017)\citenamefont {Doi},
  \citenamefont {Matsumoto}, \citenamefont {Inoue}, \citenamefont {Iwamoto},
  \citenamefont {Takano}, \citenamefont {Matsushita}, \citenamefont
  {Takahashi},\ and\ \citenamefont {Watanabe}}]{Doi2017}%
  \BibitemOpen
  \bibfield  {author} {\bibinfo {author} {\bibfnamefont {Y.}~\bibnamefont
  {Doi}}, \bibinfo {author} {\bibfnamefont {A.}~\bibnamefont {Matsumoto}},
  \bibinfo {author} {\bibfnamefont {T.}~\bibnamefont {Inoue}}, \bibinfo
  {author} {\bibfnamefont {T.}~\bibnamefont {Iwamoto}}, \bibinfo {author}
  {\bibfnamefont {A.}~\bibnamefont {Takano}}, \bibinfo {author} {\bibfnamefont
  {Y.}~\bibnamefont {Matsushita}}, \bibinfo {author} {\bibfnamefont
  {Y.}~\bibnamefont {Takahashi}},\ and\ \bibinfo {author} {\bibfnamefont
  {H.}~\bibnamefont {Watanabe}},\ }\bibfield  {title} {\bibinfo {title}
  {{Re-examination of terminal relaxation behavior of high-molecular-weight
  ring polystyrene melts}},\ }\href {https://doi.org/10.1007/s00397-017-1014-3}
  {\bibfield  {journal} {\bibinfo  {journal} {Rheologica Acta}\ }\textbf
  {\bibinfo {volume} {56}},\ \bibinfo {pages} {567} (\bibinfo {year}
  {2017})}\BibitemShut {NoStop}%
\bibitem [{\citenamefont {{Sergei P. Obukhov}}\ \emph
  {et~al.}(1994)\citenamefont {{Sergei P. Obukhov}}, \citenamefont
  {Rubinstein},\ and\ \citenamefont {Duke}}]{Obukhov1994}%
  \BibitemOpen
  \bibfield  {author} {\bibinfo {author} {\bibnamefont {{Sergei P. Obukhov}}},
  \bibinfo {author} {\bibfnamefont {M.}~\bibnamefont {Rubinstein}},\ and\
  \bibinfo {author} {\bibfnamefont {T.}~\bibnamefont {Duke}},\ }\bibfield
  {title} {\bibinfo {title} {{Dynamics of a Ring Polymer in a Gel}},\
  }\href@noop {} {\bibfield  {journal} {\bibinfo  {journal} {Physical Review
  Letters}\ }\textbf {\bibinfo {volume} {73}},\ \bibinfo {pages} {1919}
  (\bibinfo {year} {1994})}\BibitemShut {NoStop}%
\bibitem [{\citenamefont {Ge}\ \emph {et~al.}(2016)\citenamefont {Ge},
  \citenamefont {Panyukov},\ and\ \citenamefont {Rubinstein}}]{Ge2016}%
  \BibitemOpen
  \bibfield  {author} {\bibinfo {author} {\bibfnamefont {T.}~\bibnamefont
  {Ge}}, \bibinfo {author} {\bibfnamefont {S.}~\bibnamefont {Panyukov}},\ and\
  \bibinfo {author} {\bibfnamefont {M.}~\bibnamefont {Rubinstein}},\ }\bibfield
   {title} {\bibinfo {title} {{Self-Similar Conformations and Dynamics in
  Entangled Melts and Solutions of Nonconcatenated Ring Polymers}},\ }\href
  {https://doi.org/10.1021/acs.macromol.5b02319} {\bibfield  {journal}
  {\bibinfo  {journal} {Macromolecules}\ }\textbf {\bibinfo {volume} {49}},\
  \bibinfo {pages} {708} (\bibinfo {year} {2016})}\BibitemShut {NoStop}%
\bibitem [{\citenamefont {McKenna}\ and\ \citenamefont
  {Plazek}(1986)}]{McKenna1986a}%
  \BibitemOpen
  \bibfield  {author} {\bibinfo {author} {\bibfnamefont {G.}~\bibnamefont
  {McKenna}}\ and\ \bibinfo {author} {\bibfnamefont {D.}~\bibnamefont
  {Plazek}},\ }\bibfield  {title} {\bibinfo {title} {{Viscosity of blends of
  linear and cyclic molecules of similar molecular mass}},\ }\href
  {http://eds.a.ebscohost.com/eds/detail/detail?vid=0{\&}sid=9a98d512-170d-46c9-9993-04daf6c36659{\%}40sdc-v-sessmgr05{\&}bdata=JnNpdGU9ZWRzLWxpdmUmc2NvcGU9c2l0ZQ{\%}3D{\%}3D{\#}AN=edselc.2-52.0-0022796932{\&}db=edselc}
  {\bibfield  {journal} {\bibinfo  {journal} {Polymer Communications
  Guildford}\ }\textbf {\bibinfo {volume} {27}},\ \bibinfo {pages} {304}
  (\bibinfo {year} {1986})}\BibitemShut {NoStop}%
\bibitem [{\citenamefont {Lee}\ \emph {et~al.}(2000)\citenamefont {Lee},
  \citenamefont {Lee}, \citenamefont {Lee}, \citenamefont {Chang},\ and\
  \citenamefont {Roovers}}]{Lee2000}%
  \BibitemOpen
  \bibfield  {author} {\bibinfo {author} {\bibfnamefont {H.~H.~C.}\
  \bibnamefont {Lee}}, \bibinfo {author} {\bibfnamefont {H.~H.~C.}\
  \bibnamefont {Lee}}, \bibinfo {author} {\bibfnamefont {W.}~\bibnamefont
  {Lee}}, \bibinfo {author} {\bibfnamefont {T.}~\bibnamefont {Chang}},\ and\
  \bibinfo {author} {\bibfnamefont {J.}~\bibnamefont {Roovers}},\ }\bibfield
  {title} {\bibinfo {title} {{Fractionation of Cyclic Polystyrene from Linear
  Precursor by HPLC at the Chromatographic Critical Condition}},\ }\href
  {https://doi.org/10.1021/ma000807b} {\bibfield  {journal} {\bibinfo
  {journal} {Macromolecules}\ }\textbf {\bibinfo {volume} {33}},\ \bibinfo
  {pages} {8119} (\bibinfo {year} {2000})},\ \Eprint
  {https://arxiv.org/abs/arXiv:1011.1669v3} {arXiv:arXiv:1011.1669v3}
  \BibitemShut {NoStop}%
\bibitem [{\citenamefont {Yan}\ \emph {et~al.}(2016)\citenamefont {Yan},
  \citenamefont {Costanzo}, \citenamefont {Jeong}, \citenamefont {Chang},\ and\
  \citenamefont {Vlassopoulos}}]{Yan2016}%
  \BibitemOpen
  \bibfield  {author} {\bibinfo {author} {\bibfnamefont {Z.~C.}\ \bibnamefont
  {Yan}}, \bibinfo {author} {\bibfnamefont {S.}~\bibnamefont {Costanzo}},
  \bibinfo {author} {\bibfnamefont {Y.}~\bibnamefont {Jeong}}, \bibinfo
  {author} {\bibfnamefont {T.}~\bibnamefont {Chang}},\ and\ \bibinfo {author}
  {\bibfnamefont {D.}~\bibnamefont {Vlassopoulos}},\ }\bibfield  {title}
  {\bibinfo {title} {{Linear and Nonlinear Shear Rheology of a Marginally
  Entangled Ring Polymer}},\ }\href
  {https://doi.org/10.1021/acs.macromol.5b02651} {\bibfield  {journal}
  {\bibinfo  {journal} {Macromolecules}\ }\textbf {\bibinfo {volume} {49}},\
  \bibinfo {pages} {1444} (\bibinfo {year} {2016})}\BibitemShut {NoStop}%
\bibitem [{\citenamefont {Huang}\ \emph {et~al.}(2019)\citenamefont {Huang},
  \citenamefont {Ahn}, \citenamefont {Parisi}, \citenamefont {Chang},
  \citenamefont {Hassager}, \citenamefont {Panyukov}, \citenamefont
  {Rubinstein},\ and\ \citenamefont {Vlassopoulos}}]{Huang2019a}%
  \BibitemOpen
  \bibfield  {author} {\bibinfo {author} {\bibfnamefont {Q.}~\bibnamefont
  {Huang}}, \bibinfo {author} {\bibfnamefont {J.}~\bibnamefont {Ahn}}, \bibinfo
  {author} {\bibfnamefont {D.}~\bibnamefont {Parisi}}, \bibinfo {author}
  {\bibfnamefont {T.}~\bibnamefont {Chang}}, \bibinfo {author} {\bibfnamefont
  {O.}~\bibnamefont {Hassager}}, \bibinfo {author} {\bibfnamefont
  {S.}~\bibnamefont {Panyukov}}, \bibinfo {author} {\bibfnamefont
  {M.}~\bibnamefont {Rubinstein}},\ and\ \bibinfo {author} {\bibfnamefont
  {D.}~\bibnamefont {Vlassopoulos}},\ }\bibfield  {title} {\bibinfo {title}
  {{Unexpected Stretching of Entangled Ring Macromolecules}},\ }\href
  {https://doi.org/10.1103/PhysRevLett.122.208001} {\bibfield  {journal}
  {\bibinfo  {journal} {Physical Review Letters}\ }\textbf {\bibinfo {volume}
  {122}},\ \bibinfo {pages} {208001} (\bibinfo {year} {2019})}\BibitemShut
  {NoStop}%
\bibitem [{\citenamefont {Borger}\ \emph {et~al.}(2020)\citenamefont {Borger},
  \citenamefont {Wang}, \citenamefont {O'Connor}, \citenamefont {Ge},
  \citenamefont {Grest}, \citenamefont {Jensen}, \citenamefont {Ahn},
  \citenamefont {Chang}, \citenamefont {Hassager}, \citenamefont {Mortensen},
  \citenamefont {Vlassopoulos},\ and\ \citenamefont {Huang}}]{Borger2020}%
  \BibitemOpen
  \bibfield  {author} {\bibinfo {author} {\bibfnamefont {A.}~\bibnamefont
  {Borger}}, \bibinfo {author} {\bibfnamefont {W.}~\bibnamefont {Wang}},
  \bibinfo {author} {\bibfnamefont {T.~C.}\ \bibnamefont {O'Connor}}, \bibinfo
  {author} {\bibfnamefont {T.}~\bibnamefont {Ge}}, \bibinfo {author}
  {\bibfnamefont {G.~S.}\ \bibnamefont {Grest}}, \bibinfo {author}
  {\bibfnamefont {G.~V.}\ \bibnamefont {Jensen}}, \bibinfo {author}
  {\bibfnamefont {J.}~\bibnamefont {Ahn}}, \bibinfo {author} {\bibfnamefont
  {T.}~\bibnamefont {Chang}}, \bibinfo {author} {\bibfnamefont
  {O.}~\bibnamefont {Hassager}}, \bibinfo {author} {\bibfnamefont
  {K.}~\bibnamefont {Mortensen}}, \bibinfo {author} {\bibfnamefont
  {D.}~\bibnamefont {Vlassopoulos}},\ and\ \bibinfo {author} {\bibfnamefont
  {Q.}~\bibnamefont {Huang}},\ }\bibfield  {title} {\bibinfo {title}
  {{Threading–Unthreading Transition of Linear-Ring Polymer Blends in
  Extensional Flow}},\ }\href {https://doi.org/10.1021/acsmacrolett.0c00607}
  {\bibfield  {journal} {\bibinfo  {journal} {ACS Macro Letters}\ }\textbf
  {\bibinfo {volume} {9}},\ \bibinfo {pages} {1452} (\bibinfo {year}
  {2020})}\BibitemShut {NoStop}%
\bibitem [{\citenamefont {Doi}\ \emph {et~al.}(2015)\citenamefont {Doi},
  \citenamefont {Matsubara}, \citenamefont {Ohta}, \citenamefont {Nakano},
  \citenamefont {Kawaguchi}, \citenamefont {Takahashi}, \citenamefont
  {Takano},\ and\ \citenamefont {Matsushita}}]{Doi2015}%
  \BibitemOpen
  \bibfield  {author} {\bibinfo {author} {\bibfnamefont {Y.}~\bibnamefont
  {Doi}}, \bibinfo {author} {\bibfnamefont {K.}~\bibnamefont {Matsubara}},
  \bibinfo {author} {\bibfnamefont {Y.}~\bibnamefont {Ohta}}, \bibinfo {author}
  {\bibfnamefont {T.}~\bibnamefont {Nakano}}, \bibinfo {author} {\bibfnamefont
  {D.}~\bibnamefont {Kawaguchi}}, \bibinfo {author} {\bibfnamefont
  {Y.}~\bibnamefont {Takahashi}}, \bibinfo {author} {\bibfnamefont
  {A.}~\bibnamefont {Takano}},\ and\ \bibinfo {author} {\bibfnamefont
  {Y.}~\bibnamefont {Matsushita}},\ }\bibfield  {title} {\bibinfo {title}
  {{Melt rheology of ring polystyrenes with ultrahigh purity}},\ }\href
  {https://doi.org/10.1021/acs.macromol.5b00076} {\bibfield  {journal}
  {\bibinfo  {journal} {Macromolecules}\ }\textbf {\bibinfo {volume} {48}},\
  \bibinfo {pages} {3140} (\bibinfo {year} {2015})}\BibitemShut {NoStop}%
\bibitem [{\citenamefont {Molnar}\ \emph {et~al.}(2021)\citenamefont {Molnar},
  \citenamefont {Helfer}, \citenamefont {Kaszas}, \citenamefont {Krisch},
  \citenamefont {Chen}, \citenamefont {McKenna}, \citenamefont {Kornfield},\
  and\ \citenamefont {Puskas}}]{mckenna2021}%
  \BibitemOpen
  \bibfield  {author} {\bibinfo {author} {\bibfnamefont {K.}~\bibnamefont
  {Molnar}}, \bibinfo {author} {\bibfnamefont {C.~A.}\ \bibnamefont {Helfer}},
  \bibinfo {author} {\bibfnamefont {G.}~\bibnamefont {Kaszas}}, \bibinfo
  {author} {\bibfnamefont {E.}~\bibnamefont {Krisch}}, \bibinfo {author}
  {\bibfnamefont {D.}~\bibnamefont {Chen}}, \bibinfo {author} {\bibfnamefont
  {G.~B.}\ \bibnamefont {McKenna}}, \bibinfo {author} {\bibfnamefont {J.~A.}\
  \bibnamefont {Kornfield}},\ and\ \bibinfo {author} {\bibfnamefont {J.~E.}\
  \bibnamefont {Puskas}},\ }\bibfield  {title} {\bibinfo {title} {Liquid
  chromatography at critical conditions ({LCCC}): Capabilities and limitations
  for polymer analysis},\ }\href@noop {} {\bibfield  {journal} {\bibinfo
  {journal} {Journal of Molecular Liquids}\ }\textbf {\bibinfo {volume} {322}}
  (\bibinfo {year} {2021})}\BibitemShut {NoStop}%
\bibitem [{\citenamefont {Chapman}\ \emph {et~al.}(2012)\citenamefont
  {Chapman}, \citenamefont {Shanbhag}, \citenamefont {Smith},\ and\
  \citenamefont {Robertson-Anderson}}]{Chapman2012a}%
  \BibitemOpen
  \bibfield  {author} {\bibinfo {author} {\bibfnamefont {C.~D.}\ \bibnamefont
  {Chapman}}, \bibinfo {author} {\bibfnamefont {S.}~\bibnamefont {Shanbhag}},
  \bibinfo {author} {\bibfnamefont {D.~E.}\ \bibnamefont {Smith}},\ and\
  \bibinfo {author} {\bibfnamefont {R.~M.}\ \bibnamefont
  {Robertson-Anderson}},\ }\bibfield  {title} {\bibinfo {title} {{Complex
  effects of molecular topology on diffusion in entangled biopolymer blends}},\
  }\href {https://doi.org/10.1039/c2sm26279g} {\bibfield  {journal} {\bibinfo
  {journal} {Soft Matter}\ }\textbf {\bibinfo {volume} {8}},\ \bibinfo {pages}
  {9177} (\bibinfo {year} {2012})}\BibitemShut {NoStop}%
\bibitem [{\citenamefont {Iyer}\ \emph {et~al.}(2007)\citenamefont {Iyer},
  \citenamefont {Lele},\ and\ \citenamefont {Shanbhag}}]{Iyer2007}%
  \BibitemOpen
  \bibfield  {author} {\bibinfo {author} {\bibfnamefont {B.~V.~S.}\
  \bibnamefont {Iyer}}, \bibinfo {author} {\bibfnamefont {A.~K.}\ \bibnamefont
  {Lele}},\ and\ \bibinfo {author} {\bibfnamefont {S.}~\bibnamefont
  {Shanbhag}},\ }\bibfield  {title} {\bibinfo {title} {{What Is the Size of a
  Ring Polymer in a Ring-Linear Blend?}},\ }\href@noop {} {\bibfield  {journal}
  {\bibinfo  {journal} {Macromelecules}\ }\textbf {\bibinfo {volume} {40}},\
  \bibinfo {pages} {5995} (\bibinfo {year} {2007})}\BibitemShut {NoStop}%
\bibitem [{\citenamefont {Halverson}\ \emph {et~al.}(2012)\citenamefont
  {Halverson}, \citenamefont {Grest}, \citenamefont {Grosberg},\ and\
  \citenamefont {Kremer}}]{Halverson2012}%
  \BibitemOpen
  \bibfield  {author} {\bibinfo {author} {\bibfnamefont {J.~D.}\ \bibnamefont
  {Halverson}}, \bibinfo {author} {\bibfnamefont {G.~S.}\ \bibnamefont
  {Grest}}, \bibinfo {author} {\bibfnamefont {A.~Y.}\ \bibnamefont
  {Grosberg}},\ and\ \bibinfo {author} {\bibfnamefont {K.}~\bibnamefont
  {Kremer}},\ }\bibfield  {title} {\bibinfo {title} {{Rheology of ring polymer
  melts: From linear contaminants to ring-linear blends}},\ }\href
  {https://doi.org/10.1103/PhysRevLett.108.038301} {\bibfield  {journal}
  {\bibinfo  {journal} {Physical Review Letters}\ }\textbf {\bibinfo {volume}
  {108}},\ \bibinfo {pages} {038301} (\bibinfo {year} {2012})},\ \Eprint
  {https://arxiv.org/abs/1112.3519} {arXiv:1112.3519} \BibitemShut {NoStop}%
\bibitem [{\citenamefont {Gartner}\ \emph {et~al.}(2019)\citenamefont
  {Gartner}, \citenamefont {Haque}, \citenamefont {Gomi}, \citenamefont
  {Grayson}, \citenamefont {Hore},\ and\ \citenamefont
  {Jayaraman}}]{Gartner2019}%
  \BibitemOpen
  \bibfield  {author} {\bibinfo {author} {\bibfnamefont {T.~E.}\ \bibnamefont
  {Gartner}}, \bibinfo {author} {\bibfnamefont {F.~M.}\ \bibnamefont {Haque}},
  \bibinfo {author} {\bibfnamefont {A.~M.}\ \bibnamefont {Gomi}}, \bibinfo
  {author} {\bibfnamefont {S.~M.}\ \bibnamefont {Grayson}}, \bibinfo {author}
  {\bibfnamefont {M.~J.~A.}\ \bibnamefont {Hore}},\ and\ \bibinfo {author}
  {\bibfnamefont {A.}~\bibnamefont {Jayaraman}},\ }\bibfield  {title} {\bibinfo
  {title} {{Scaling Exponent and Effective Interactions in Linear and Cyclic
  Polymer Solutions: Theory, Simulations, and Experiments}},\ }\href
  {https://doi.org/10.1021/acs.macromol.9b00600} {\bibfield  {journal}
  {\bibinfo  {journal} {Macromolecules}\ }\textbf {\bibinfo {volume} {52}},\
  \bibinfo {pages} {4579} (\bibinfo {year} {2019})}\BibitemShut {NoStop}%
\bibitem [{\citenamefont {Robertson}\ and\ \citenamefont
  {Smith}(2007{\natexlab{a}})}]{Robertson2007c}%
  \BibitemOpen
  \bibfield  {author} {\bibinfo {author} {\bibfnamefont {R.~M.}\ \bibnamefont
  {Robertson}}\ and\ \bibinfo {author} {\bibfnamefont {D.~E.}\ \bibnamefont
  {Smith}},\ }\bibfield  {title} {\bibinfo {title} {{Strong effects of
  molecular topology on diffusion of entangled DNA molecules.}},\ }\href
  {https://doi.org/10.1073/pnas.0700137104} {\bibfield  {journal} {\bibinfo
  {journal} {Proceedings of the National Academy of Sciences of the United
  States of America}\ }\textbf {\bibinfo {volume} {104}},\ \bibinfo {pages}
  {4824} (\bibinfo {year} {2007}{\natexlab{a}})}\BibitemShut {NoStop}%
\bibitem [{\citenamefont {Robertson}\ and\ \citenamefont
  {Smith}(2007{\natexlab{b}})}]{Robertson2007a}%
  \BibitemOpen
  \bibfield  {author} {\bibinfo {author} {\bibfnamefont {R.~M.}\ \bibnamefont
  {Robertson}}\ and\ \bibinfo {author} {\bibfnamefont {D.~E.}\ \bibnamefont
  {Smith}},\ }\bibfield  {title} {\bibinfo {title} {{Self-diffusion of
  entangled linear and circular DNA molecules: Dependence on length and
  concentration}},\ }\href {https://doi.org/10.1021/ma070051h} {\bibfield
  {journal} {\bibinfo  {journal} {Macromolecules}\ }\textbf {\bibinfo {volume}
  {40}},\ \bibinfo {pages} {3373} (\bibinfo {year}
  {2007}{\natexlab{b}})}\BibitemShut {NoStop}%
\bibitem [{\citenamefont {Subramanian}\ and\ \citenamefont
  {Shanbhag}(2008)}]{Subramanian2008}%
  \BibitemOpen
  \bibfield  {author} {\bibinfo {author} {\bibfnamefont {G.}~\bibnamefont
  {Subramanian}}\ and\ \bibinfo {author} {\bibfnamefont {S.}~\bibnamefont
  {Shanbhag}},\ }\bibfield  {title} {\bibinfo {title} {{Self-diffusion in
  binary blends of cyclic and linear polymers}},\ }\href
  {https://doi.org/10.1021/ma801232j} {\bibfield  {journal} {\bibinfo
  {journal} {Macromolecules}\ }\textbf {\bibinfo {volume} {41}},\ \bibinfo
  {pages} {7239} (\bibinfo {year} {2008})}\BibitemShut {NoStop}%
\bibitem [{\citenamefont {Habuchi}\ \emph {et~al.}(2010)\citenamefont
  {Habuchi}, \citenamefont {Satoh}, \citenamefont {Yamamoto}, \citenamefont
  {Tezuka},\ and\ \citenamefont {Vacha}}]{Habuchi2010}%
  \BibitemOpen
  \bibfield  {author} {\bibinfo {author} {\bibfnamefont {S.}~\bibnamefont
  {Habuchi}}, \bibinfo {author} {\bibfnamefont {N.}~\bibnamefont {Satoh}},
  \bibinfo {author} {\bibfnamefont {T.}~\bibnamefont {Yamamoto}}, \bibinfo
  {author} {\bibfnamefont {Y.}~\bibnamefont {Tezuka}},\ and\ \bibinfo {author}
  {\bibfnamefont {M.}~\bibnamefont {Vacha}},\ }\bibfield  {title} {\bibinfo
  {title} {{Multimode diffusion of ring polymer molecules revealed by a
  single-molecule study}},\ }\href {https://doi.org/10.1002/anie.200904394}
  {\bibfield  {journal} {\bibinfo  {journal} {Angewandte Chemie - International
  Edition}\ }\textbf {\bibinfo {volume} {49}},\ \bibinfo {pages} {1418}
  (\bibinfo {year} {2010})}\BibitemShut {NoStop}%
\bibitem [{\citenamefont {Graessley}(1982)}]{Graessley1982}%
  \BibitemOpen
  \bibfield  {author} {\bibinfo {author} {\bibfnamefont {W.}~\bibnamefont
  {Graessley}},\ }\bibfield  {title} {\bibinfo {title} {{Entangled linear,
  branched and network polymer systems - Molecular theories}},\ }in\ \href
  {https://doi.org/10.1007/bfb0038532} {\emph {\bibinfo {booktitle} {Synthesis
  and Degradation Rheology and Extrusion}}}\ (\bibinfo  {publisher}
  {Springer-Verlag Berlin Heidelberg},\ \bibinfo {year} {1982})\ pp.\ \bibinfo
  {pages} {67--117}\BibitemShut {NoStop}%
\bibitem [{\citenamefont {Klein}(1986)}]{Klein1986}%
  \BibitemOpen
  \bibfield  {author} {\bibinfo {author} {\bibfnamefont {J.}~\bibnamefont
  {Klein}},\ }\bibfield  {title} {\bibinfo {title} {{Dynamics of Entangled
  Linear, Branched, and Cyclic Polymers}},\ }\href
  {https://doi.org/10.1021/ma00155a018} {\bibfield  {journal} {\bibinfo
  {journal} {Macromolecules}\ }\textbf {\bibinfo {volume} {19}},\ \bibinfo
  {pages} {105} (\bibinfo {year} {1986})}\BibitemShut {NoStop}%
\bibitem [{\citenamefont {Mills}\ \emph {et~al.}(1987)\citenamefont {Mills},
  \citenamefont {Mayer}, \citenamefont {Kramer}, \citenamefont {Hadziioannou},
  \citenamefont {Lutz}, \citenamefont {Strazielle}, \citenamefont {Rempp},\
  and\ \citenamefont {Kovacs}}]{Mills1987}%
  \BibitemOpen
  \bibfield  {author} {\bibinfo {author} {\bibfnamefont {P.~J.}\ \bibnamefont
  {Mills}}, \bibinfo {author} {\bibfnamefont {J.~W.}\ \bibnamefont {Mayer}},
  \bibinfo {author} {\bibfnamefont {E.~J.}\ \bibnamefont {Kramer}}, \bibinfo
  {author} {\bibfnamefont {G.}~\bibnamefont {Hadziioannou}}, \bibinfo {author}
  {\bibfnamefont {P.}~\bibnamefont {Lutz}}, \bibinfo {author} {\bibfnamefont
  {C.}~\bibnamefont {Strazielle}}, \bibinfo {author} {\bibfnamefont
  {P.}~\bibnamefont {Rempp}},\ and\ \bibinfo {author} {\bibfnamefont {a.~J.}\
  \bibnamefont {Kovacs}},\ }\bibfield  {title} {\bibinfo {title} {{Diffusion of
  polymer rings in linear polymer matrices}},\ }\href
  {https://doi.org/10.1021/ma00169a008} {\bibfield  {journal} {\bibinfo
  {journal} {Macromolecules}\ }\textbf {\bibinfo {volume} {20}},\ \bibinfo
  {pages} {513} (\bibinfo {year} {1987})}\BibitemShut {NoStop}%
\bibitem [{\citenamefont {Yang}\ \emph {et~al.}(2010)\citenamefont {Yang},
  \citenamefont {Sun}, \citenamefont {Fu}, \citenamefont {An},\ and\
  \citenamefont {Wang}}]{Yang2010}%
  \BibitemOpen
  \bibfield  {author} {\bibinfo {author} {\bibfnamefont {Y.-B.}\ \bibnamefont
  {Yang}}, \bibinfo {author} {\bibfnamefont {Z.-Y.}\ \bibnamefont {Sun}},
  \bibinfo {author} {\bibfnamefont {C.-L.}\ \bibnamefont {Fu}}, \bibinfo
  {author} {\bibfnamefont {L.-J.}\ \bibnamefont {An}},\ and\ \bibinfo {author}
  {\bibfnamefont {Z.-G.}\ \bibnamefont {Wang}},\ }\bibfield  {title} {\bibinfo
  {title} {{Monte Carlo simulation of a single ring among linear chains:
  structural and dynamic heterogeneity.}},\ }\href
  {https://doi.org/10.1063/1.3466921} {\bibfield  {journal} {\bibinfo
  {journal} {Journal of Chemical Physics}\ }\textbf {\bibinfo {volume} {133}},\
  \bibinfo {pages} {064901} (\bibinfo {year} {2010})}\BibitemShut {NoStop}%
\bibitem [{\citenamefont {Tsalikis}\ \emph {et~al.}(2016)\citenamefont
  {Tsalikis}, \citenamefont {Mavrantzas},\ and\ \citenamefont
  {Vlassopoulos}}]{Tsalikis2016}%
  \BibitemOpen
  \bibfield  {author} {\bibinfo {author} {\bibfnamefont {D.~G.}\ \bibnamefont
  {Tsalikis}}, \bibinfo {author} {\bibfnamefont {V.~G.}\ \bibnamefont
  {Mavrantzas}},\ and\ \bibinfo {author} {\bibfnamefont {D.}~\bibnamefont
  {Vlassopoulos}},\ }\bibfield  {title} {\bibinfo {title} {{Analysis of Slow
  Modes in Ring Polymers: Threading of Rings Controls Long-Time Relaxation}},\
  }\href {https://doi.org/10.1021/acsmacrolett.6b00259} {\bibfield  {journal}
  {\bibinfo  {journal} {ACS Macro Letters}\ }\textbf {\bibinfo {volume} {5}},\
  \bibinfo {pages} {755} (\bibinfo {year} {2016})}\BibitemShut {NoStop}%
\bibitem [{\citenamefont {Kruteva}\ \emph {et~al.}(2017)\citenamefont
  {Kruteva}, \citenamefont {Allgaier},\ and\ \citenamefont
  {Richter}}]{Kruteva2017}%
  \BibitemOpen
  \bibfield  {author} {\bibinfo {author} {\bibfnamefont {M.}~\bibnamefont
  {Kruteva}}, \bibinfo {author} {\bibfnamefont {J.}~\bibnamefont {Allgaier}},\
  and\ \bibinfo {author} {\bibfnamefont {D.}~\bibnamefont {Richter}},\
  }\bibfield  {title} {\bibinfo {title} {{Direct observation of two distinct
  diffusive modes for polymer rings in linear polymer matrices by pulsed field
  gradient (PFG) NMR}},\ }\href {https://doi.org/10.1021/acs.macromol.7b01850}
  {\bibfield  {journal} {\bibinfo  {journal} {Macromolecules}\ }\textbf
  {\bibinfo {volume} {50}},\ \bibinfo {pages} {9482} (\bibinfo {year}
  {2017})}\BibitemShut {NoStop}%
\bibitem [{\citenamefont {O’Connor}\ \emph {et~al.}(2020)\citenamefont
  {O’Connor}, \citenamefont {Ge}, \citenamefont {Rubinstein},\ and\
  \citenamefont {Grest}}]{o2020topological}%
  \BibitemOpen
  \bibfield  {author} {\bibinfo {author} {\bibfnamefont {T.~C.}\ \bibnamefont
  {O’Connor}}, \bibinfo {author} {\bibfnamefont {T.}~\bibnamefont {Ge}},
  \bibinfo {author} {\bibfnamefont {M.}~\bibnamefont {Rubinstein}},\ and\
  \bibinfo {author} {\bibfnamefont {G.~S.}\ \bibnamefont {Grest}},\ }\bibfield
  {title} {\bibinfo {title} {Topological linking drives anomalous thickening of
  ring polymers in weak extensional flows},\ }\href@noop {} {\bibfield
  {journal} {\bibinfo  {journal} {Physical Review Letters}\ }\textbf {\bibinfo
  {volume} {124}},\ \bibinfo {pages} {027801} (\bibinfo {year}
  {2020})}\BibitemShut {NoStop}%
\bibitem [{\citenamefont {Schroeder}(2018)}]{Schroeder2018}%
  \BibitemOpen
  \bibfield  {author} {\bibinfo {author} {\bibfnamefont {C.~M.}\ \bibnamefont
  {Schroeder}},\ }\bibfield  {title} {\bibinfo {title} {{Single Polymer
  Dynamics for Molecular Rheology}},\ }\href
  {https://doi.org/10.1122/1.5013246} {\bibfield  {journal} {\bibinfo
  {journal} {Journal of Rheology}\ }\textbf {\bibinfo {volume} {62}},\ \bibinfo
  {pages} {371} (\bibinfo {year} {2018})}\BibitemShut {NoStop}%
\bibitem [{\citenamefont {Perkins}\ \emph {et~al.}(1997)\citenamefont
  {Perkins}, \citenamefont {Smith},\ and\ \citenamefont {Chu}}]{Perkins1997}%
  \BibitemOpen
  \bibfield  {author} {\bibinfo {author} {\bibfnamefont {T.~T.}\ \bibnamefont
  {Perkins}}, \bibinfo {author} {\bibfnamefont {D.~E.}\ \bibnamefont {Smith}},\
  and\ \bibinfo {author} {\bibfnamefont {S.}~\bibnamefont {Chu}},\ }\bibfield
  {title} {\bibinfo {title} {{Single polymer dynamics in an elongational
  flow}},\ }\href {https://doi.org/10.1126/science.276.5321.2016} {\bibfield
  {journal} {\bibinfo  {journal} {Science}\ }\textbf {\bibinfo {volume}
  {276}},\ \bibinfo {pages} {2016} (\bibinfo {year} {1997})}\BibitemShut
  {NoStop}%
\bibitem [{\citenamefont {Smith}\ \emph {et~al.}(1999)\citenamefont {Smith},
  \citenamefont {Babcock},\ and\ \citenamefont {Chu}}]{Smith1999}%
  \BibitemOpen
  \bibfield  {author} {\bibinfo {author} {\bibfnamefont {D.~E.}\ \bibnamefont
  {Smith}}, \bibinfo {author} {\bibfnamefont {H.~P.}\ \bibnamefont {Babcock}},\
  and\ \bibinfo {author} {\bibfnamefont {S.}~\bibnamefont {Chu}},\ }\bibfield
  {title} {\bibinfo {title} {{Single-polymer dynamics in steady shear flow}},\
  }\href {https://doi.org/10.1126/science.283.5408.1724} {\bibfield  {journal}
  {\bibinfo  {journal} {Science}\ }\textbf {\bibinfo {volume} {283}},\ \bibinfo
  {pages} {1724} (\bibinfo {year} {1999})}\BibitemShut {NoStop}%
\bibitem [{\citenamefont {Soh}\ \emph {et~al.}(2018)\citenamefont {Soh},
  \citenamefont {Narsimhan}, \citenamefont {Klotz},\ and\ \citenamefont
  {Doyle}}]{Soh2018}%
  \BibitemOpen
  \bibfield  {author} {\bibinfo {author} {\bibfnamefont {B.~W.}\ \bibnamefont
  {Soh}}, \bibinfo {author} {\bibfnamefont {V.}~\bibnamefont {Narsimhan}},
  \bibinfo {author} {\bibfnamefont {A.~R.}\ \bibnamefont {Klotz}},\ and\
  \bibinfo {author} {\bibfnamefont {P.~S.}\ \bibnamefont {Doyle}},\ }\bibfield
  {title} {\bibinfo {title} {{Knots modify the coil-stretch transition in
  linear DNA polymers}},\ }\href {https://doi.org/10.1039/c7sm02195j}
  {\bibfield  {journal} {\bibinfo  {journal} {Soft Matter}\ }\textbf {\bibinfo
  {volume} {14}},\ \bibinfo {pages} {1689} (\bibinfo {year}
  {2018})}\BibitemShut {NoStop}%
\bibitem [{\citenamefont {Zhou}\ and\ \citenamefont
  {Schroeder}(2016{\natexlab{a}})}]{Zhou2016}%
  \BibitemOpen
  \bibfield  {author} {\bibinfo {author} {\bibfnamefont {Y.}~\bibnamefont
  {Zhou}}\ and\ \bibinfo {author} {\bibfnamefont {C.~M.}\ \bibnamefont
  {Schroeder}},\ }\bibfield  {title} {\bibinfo {title} {{Single polymer
  dynamics under large amplitude oscillatory extension}},\ }\href
  {https://doi.org/10.1103/PhysRevFluids.1.053301} {\bibfield  {journal}
  {\bibinfo  {journal} {Physical Review Fluids}\ }\textbf {\bibinfo {volume}
  {1}},\ \bibinfo {pages} {053301} (\bibinfo {year}
  {2016}{\natexlab{a}})}\BibitemShut {NoStop}%
\bibitem [{\citenamefont {Zhou}\ and\ \citenamefont
  {Schroeder}(2016{\natexlab{b}})}]{Zhou2016b}%
  \BibitemOpen
  \bibfield  {author} {\bibinfo {author} {\bibfnamefont {Y.}~\bibnamefont
  {Zhou}}\ and\ \bibinfo {author} {\bibfnamefont {C.~M.}\ \bibnamefont
  {Schroeder}},\ }\bibfield  {title} {\bibinfo {title} {{Transient and Average
  Unsteady Dynamics of Single Polymers in Large-Amplitude Oscillatory
  Extension}},\ }\href {https://doi.org/10.1021/acs.macromol.6b01606}
  {\bibfield  {journal} {\bibinfo  {journal} {Macromolecules}\ }\textbf
  {\bibinfo {volume} {49}},\ \bibinfo {pages} {8018} (\bibinfo {year}
  {2016}{\natexlab{b}})}\BibitemShut {NoStop}%
\bibitem [{\citenamefont {Hsiao}\ \emph {et~al.}(2017)\citenamefont {Hsiao},
  \citenamefont {Samsal}, \citenamefont {Prakash},\ and\ \citenamefont
  {Schroeder}}]{Hsiao2017}%
  \BibitemOpen
  \bibfield  {author} {\bibinfo {author} {\bibfnamefont {K.-W.}\ \bibnamefont
  {Hsiao}}, \bibinfo {author} {\bibfnamefont {C.}~\bibnamefont {Samsal}},
  \bibinfo {author} {\bibfnamefont {J.~R.}\ \bibnamefont {Prakash}},\ and\
  \bibinfo {author} {\bibfnamefont {C.~M.}\ \bibnamefont {Schroeder}},\
  }\bibfield  {title} {\bibinfo {title} {{Direct observation of DNA dynamics in
  semi-dilute solutions in extensional flow}},\ }\href
  {https://doi.org/10.1122/1.4972236} {\bibfield  {journal} {\bibinfo
  {journal} {Journal of Rheology}\ }\textbf {\bibinfo {volume} {61}},\ \bibinfo
  {pages} {151} (\bibinfo {year} {2017})},\ \Eprint
  {https://arxiv.org/abs/1604.06754} {arXiv:1604.06754} \BibitemShut {NoStop}%
\bibitem [{\citenamefont {Samsal}\ \emph {et~al.}(2017)\citenamefont {Samsal},
  \citenamefont {Hsiao}, \citenamefont {Schroeder},\ and\ \citenamefont
  {Prakash}}]{Samsal2017}%
  \BibitemOpen
  \bibfield  {author} {\bibinfo {author} {\bibfnamefont {C.}~\bibnamefont
  {Samsal}}, \bibinfo {author} {\bibfnamefont {K.-W.}\ \bibnamefont {Hsiao}},
  \bibinfo {author} {\bibfnamefont {C.~M.}\ \bibnamefont {Schroeder}},\ and\
  \bibinfo {author} {\bibfnamefont {J.~R.}\ \bibnamefont {Prakash}},\
  }\bibfield  {title} {\bibinfo {title} {{Parameter-Free Prediction of DNA
  dynamics in Planar Extensional Flow of Semidilute Solutions}},\ }\href
  {https://doi.org/10.1122/1.4972237} {\bibfield  {journal} {\bibinfo
  {journal} {Journal of Rheology}\ }\textbf {\bibinfo {volume} {61}},\ \bibinfo
  {pages} {169} (\bibinfo {year} {2017})}\BibitemShut {NoStop}%
\bibitem [{\citenamefont {Young}\ and\ \citenamefont
  {Sing}(2019{\natexlab{a}})}]{Young2019a}%
  \BibitemOpen
  \bibfield  {author} {\bibinfo {author} {\bibfnamefont {C.~D.}\ \bibnamefont
  {Young}}\ and\ \bibinfo {author} {\bibfnamefont {C.~E.}\ \bibnamefont
  {Sing}},\ }\bibfield  {title} {\bibinfo {title} {{Simulation of semidilute
  polymer solutions in planar extensional flow via conformationally averaged
  Brownian noise}},\ }\href {https://doi.org/10.1063/1.5122811} {\bibfield
  {journal} {\bibinfo  {journal} {Journal of Chemical Physics}\ }\textbf
  {\bibinfo {volume} {151}},\ \bibinfo {pages} {124907} (\bibinfo {year}
  {2019}{\natexlab{a}})}\BibitemShut {NoStop}%
\bibitem [{\citenamefont {Zhou}\ and\ \citenamefont
  {Schroeder}(2018)}]{Zhou2018}%
  \BibitemOpen
  \bibfield  {author} {\bibinfo {author} {\bibfnamefont {Y.}~\bibnamefont
  {Zhou}}\ and\ \bibinfo {author} {\bibfnamefont {C.~M.}\ \bibnamefont
  {Schroeder}},\ }\bibfield  {title} {\bibinfo {title} {{Dynamically
  Heterogeneous Relaxation of Entangled Polymer Chains}},\ }\href
  {https://doi.org/10.1103/PhysRevLett.120.267801} {\bibfield  {journal}
  {\bibinfo  {journal} {Physical Review Letters}\ }\textbf {\bibinfo {volume}
  {120}},\ \bibinfo {pages} {267801} (\bibinfo {year} {2018})}\BibitemShut
  {NoStop}%
\bibitem [{\citenamefont {Li}\ \emph {et~al.}(2015)\citenamefont {Li},
  \citenamefont {Hsiao}, \citenamefont {Brockman}, \citenamefont {Yates},
  \citenamefont {Robertson-Anderson}, \citenamefont {Kornfield}, \citenamefont
  {{San Francisco}}, \citenamefont {Schroeder},\ and\ \citenamefont
  {McKenna}}]{Li2015}%
  \BibitemOpen
  \bibfield  {author} {\bibinfo {author} {\bibfnamefont {Y.}~\bibnamefont
  {Li}}, \bibinfo {author} {\bibfnamefont {K.-W.}\ \bibnamefont {Hsiao}},
  \bibinfo {author} {\bibfnamefont {C.~A.}\ \bibnamefont {Brockman}}, \bibinfo
  {author} {\bibfnamefont {D.~Y.}\ \bibnamefont {Yates}}, \bibinfo {author}
  {\bibfnamefont {R.~M.}\ \bibnamefont {Robertson-Anderson}}, \bibinfo {author}
  {\bibfnamefont {J.~A.}\ \bibnamefont {Kornfield}}, \bibinfo {author}
  {\bibfnamefont {M.~J.}\ \bibnamefont {{San Francisco}}}, \bibinfo {author}
  {\bibfnamefont {C.~M.}\ \bibnamefont {Schroeder}},\ and\ \bibinfo {author}
  {\bibfnamefont {G.~B.}\ \bibnamefont {McKenna}},\ }\bibfield  {title}
  {\bibinfo {title} {{When ends meet: Circular DNA stretches differently in
  elongational flows}},\ }\href {https://doi.org/10.1021/acs.macromol.5b01374}
  {\bibfield  {journal} {\bibinfo  {journal} {Macromolecules}\ }\textbf
  {\bibinfo {volume} {48}},\ \bibinfo {pages} {5997} (\bibinfo {year}
  {2015})}\BibitemShut {NoStop}%
\bibitem [{\citenamefont {Hsiao}\ \emph {et~al.}(2016)\citenamefont {Hsiao},
  \citenamefont {Schroeder},\ and\ \citenamefont {Sing}}]{Hsiao2016a}%
  \BibitemOpen
  \bibfield  {author} {\bibinfo {author} {\bibfnamefont {K.-W.}\ \bibnamefont
  {Hsiao}}, \bibinfo {author} {\bibfnamefont {C.~M.}\ \bibnamefont
  {Schroeder}},\ and\ \bibinfo {author} {\bibfnamefont {C.~E.}\ \bibnamefont
  {Sing}},\ }\bibfield  {title} {\bibinfo {title} {{Ring Polymer Dynamics Are
  Governed by a Coupling between Architecture and Hydrodynamic Interactions}},\
  }\href {https://doi.org/10.1021/acs.macromol.5b02357} {\bibfield  {journal}
  {\bibinfo  {journal} {Macromolecules}\ }\textbf {\bibinfo {volume} {49}},\
  \bibinfo {pages} {1961} (\bibinfo {year} {2016})}\BibitemShut {NoStop}%
\bibitem [{\citenamefont {Weiss}\ \emph {et~al.}(2017)\citenamefont {Weiss},
  \citenamefont {Nikoubashman},\ and\ \citenamefont {Likos}}]{Weiss2017}%
  \BibitemOpen
  \bibfield  {author} {\bibinfo {author} {\bibfnamefont {L.~B.}\ \bibnamefont
  {Weiss}}, \bibinfo {author} {\bibfnamefont {A.}~\bibnamefont
  {Nikoubashman}},\ and\ \bibinfo {author} {\bibfnamefont {C.~N.}\ \bibnamefont
  {Likos}},\ }\bibfield  {title} {\bibinfo {title} {{Topology-Sensitive
  Microfluidic Filter for Polymers of Varying Stiffness}},\ }\href
  {https://doi.org/10.1021/acsmacrolett.7b00768} {\bibfield  {journal}
  {\bibinfo  {journal} {ACS Macro Letters}\ }\textbf {\bibinfo {volume} {6}},\
  \bibinfo {pages} {1426} (\bibinfo {year} {2017})}\BibitemShut {NoStop}%
\bibitem [{\citenamefont {Young}\ \emph {et~al.}(2019)\citenamefont {Young},
  \citenamefont {Qian}, \citenamefont {Marvin},\ and\ \citenamefont
  {Sing}}]{Young2019}%
  \BibitemOpen
  \bibfield  {author} {\bibinfo {author} {\bibfnamefont {C.~D.}\ \bibnamefont
  {Young}}, \bibinfo {author} {\bibfnamefont {J.~R.}\ \bibnamefont {Qian}},
  \bibinfo {author} {\bibfnamefont {M.}~\bibnamefont {Marvin}},\ and\ \bibinfo
  {author} {\bibfnamefont {C.~E.}\ \bibnamefont {Sing}},\ }\bibfield  {title}
  {\bibinfo {title} {{Ring polymer dynamics and tumbling-stretch transitions in
  planar mixed flows}},\ }\href {https://doi.org/10.1103/PhysRevE.99.062502}
  {\bibfield  {journal} {\bibinfo  {journal} {Physical Review E}\ }\textbf
  {\bibinfo {volume} {99}},\ \bibinfo {pages} {062502} (\bibinfo {year}
  {2019})}\BibitemShut {NoStop}%
\bibitem [{\citenamefont {Tu}\ \emph {et~al.}(2020)\citenamefont {Tu},
  \citenamefont {Lee}, \citenamefont {Robertson-anderson},\ and\ \citenamefont
  {Schroeder}}]{Tu2020}%
  \BibitemOpen
  \bibfield  {author} {\bibinfo {author} {\bibfnamefont {M.~Q.}\ \bibnamefont
  {Tu}}, \bibinfo {author} {\bibfnamefont {M.}~\bibnamefont {Lee}}, \bibinfo
  {author} {\bibfnamefont {R.~M.}\ \bibnamefont {Robertson-anderson}},\ and\
  \bibinfo {author} {\bibfnamefont {C.~M.}\ \bibnamefont {Schroeder}},\
  }\bibfield  {title} {\bibinfo {title} {{Direct Observation of Ring Polymer
  Dynamics in the Flow-Gradient Plane of Shear Flow}},\ }\href
  {https://doi.org/10.1021/acs.macromol.0c01362} {\bibfield  {journal}
  {\bibinfo  {journal} {Macromelecules}\ }\textbf {\bibinfo {volume} {53}},\
  \bibinfo {pages} {9406} (\bibinfo {year} {2020})}\BibitemShut {NoStop}%
\bibitem [{\citenamefont {Zhou}\ \emph {et~al.}(2019)\citenamefont {Zhou},
  \citenamefont {Hsiao}, \citenamefont {Regan}, \citenamefont {Kong},
  \citenamefont {McKenna}, \citenamefont {Robertson-Anderson},\ and\
  \citenamefont {Schroeder}}]{Zhou2019}%
  \BibitemOpen
  \bibfield  {author} {\bibinfo {author} {\bibfnamefont {Y.}~\bibnamefont
  {Zhou}}, \bibinfo {author} {\bibfnamefont {K.-W.}\ \bibnamefont {Hsiao}},
  \bibinfo {author} {\bibfnamefont {K.~E.}\ \bibnamefont {Regan}}, \bibinfo
  {author} {\bibfnamefont {D.}~\bibnamefont {Kong}}, \bibinfo {author}
  {\bibfnamefont {G.~B.}\ \bibnamefont {McKenna}}, \bibinfo {author}
  {\bibfnamefont {R.~M.}\ \bibnamefont {Robertson-Anderson}},\ and\ \bibinfo
  {author} {\bibfnamefont {C.~M.}\ \bibnamefont {Schroeder}},\ }\bibfield
  {title} {\bibinfo {title} {{Effect of molecular architecture on ring polymer
  dynamics in semidilute linear polymer solutions}},\ }\href
  {https://doi.org/10.1038/s41467-019-09627-7} {\bibfield  {journal} {\bibinfo
  {journal} {Nature Communications}\ }\textbf {\bibinfo {volume} {10}},\
  \bibinfo {pages} {1753} (\bibinfo {year} {2019})}\BibitemShut {NoStop}%
\bibitem [{\citenamefont {Young}\ \emph {et~al.}(2020)\citenamefont {Young},
  \citenamefont {Zhou}, \citenamefont {Schroeder},\ and\ \citenamefont
  {Sing}}]{young2020dynamics}%
  \BibitemOpen
  \bibfield  {author} {\bibinfo {author} {\bibfnamefont {C.~D.}\ \bibnamefont
  {Young}}, \bibinfo {author} {\bibfnamefont {Y.}~\bibnamefont {Zhou}},
  \bibinfo {author} {\bibfnamefont {C.~M.}\ \bibnamefont {Schroeder}},\ and\
  \bibinfo {author} {\bibfnamefont {C.~E.}\ \bibnamefont {Sing}},\ }\href@noop
  {} {\bibinfo {title} {Dynamics and rheology of semidilute solutions of
  ring-linear polymer blends in planar extensional flow}} (\bibinfo {year}
  {2020}),\ \Eprint {https://arxiv.org/abs/2011.01386} {arXiv:2011.01386
  [cond-mat.soft]} \BibitemShut {NoStop}%
\bibitem [{\citenamefont {Laib}\ \emph {et~al.}(2006)\citenamefont {Laib},
  \citenamefont {Robertson},\ and\ \citenamefont {Smith}}]{Laib2006a}%
  \BibitemOpen
  \bibfield  {author} {\bibinfo {author} {\bibfnamefont {S.}~\bibnamefont
  {Laib}}, \bibinfo {author} {\bibfnamefont {R.~M.}\ \bibnamefont
  {Robertson}},\ and\ \bibinfo {author} {\bibfnamefont {D.~E.}\ \bibnamefont
  {Smith}},\ }\bibfield  {title} {\bibinfo {title} {{Preparation and
  characterization of a set of linear DNA molecules for polymer physics and
  rheology studies}},\ }\href {https://doi.org/10.1021/ma0601464} {\bibfield
  {journal} {\bibinfo  {journal} {Macromolecules}\ }\textbf {\bibinfo {volume}
  {39}},\ \bibinfo {pages} {4115} (\bibinfo {year} {2006})}\BibitemShut
  {NoStop}%
\bibitem [{\citenamefont {Robertson}\ \emph {et~al.}(2006)\citenamefont
  {Robertson}, \citenamefont {Laib},\ and\ \citenamefont
  {Smith}}]{Robertson2006a}%
  \BibitemOpen
  \bibfield  {author} {\bibinfo {author} {\bibfnamefont {R.~M.}\ \bibnamefont
  {Robertson}}, \bibinfo {author} {\bibfnamefont {S.}~\bibnamefont {Laib}},\
  and\ \bibinfo {author} {\bibfnamefont {D.~E.}\ \bibnamefont {Smith}},\
  }\bibfield  {title} {\bibinfo {title} {{Diffusion of isolated DNA molecules:
  dependence on length and topology.}},\ }\href
  {https://doi.org/10.1073/pnas.0601903103} {\bibfield  {journal} {\bibinfo
  {journal} {Proceedings of the National Academy of Sciences of the United
  States of America}\ }\textbf {\bibinfo {volume} {103}},\ \bibinfo {pages}
  {7310} (\bibinfo {year} {2006})}\BibitemShut {NoStop}%
\bibitem [{\citenamefont {Peddireddy}\ \emph
  {et~al.}(2020{\natexlab{a}})\citenamefont {Peddireddy}, \citenamefont {Lee},
  \citenamefont {Zhou}, \citenamefont {Adalbert}, \citenamefont {Anderson},
  \citenamefont {Schroeder},\ and\ \citenamefont
  {Robertson-anderson}}]{Peddireddy2020}%
  \BibitemOpen
  \bibfield  {author} {\bibinfo {author} {\bibfnamefont {K.~R.}\ \bibnamefont
  {Peddireddy}}, \bibinfo {author} {\bibfnamefont {M.}~\bibnamefont {Lee}},
  \bibinfo {author} {\bibfnamefont {Y.}~\bibnamefont {Zhou}}, \bibinfo {author}
  {\bibfnamefont {S.}~\bibnamefont {Adalbert}}, \bibinfo {author}
  {\bibfnamefont {S.}~\bibnamefont {Anderson}}, \bibinfo {author}
  {\bibfnamefont {C.~M.}\ \bibnamefont {Schroeder}},\ and\ \bibinfo {author}
  {\bibfnamefont {R.~M.}\ \bibnamefont {Robertson-anderson}},\ }\bibfield
  {title} {\bibinfo {title} {{Unexpected entanglement dynamics in semidilute
  blends of supercoiled and ring DNA}},\ }\href
  {https://doi.org/10.1039/c9sm01767d} {\bibfield  {journal} {\bibinfo
  {journal} {Soft Matter}\ }\textbf {\bibinfo {volume} {16}},\ \bibinfo {pages}
  {152} (\bibinfo {year} {2020}{\natexlab{a}})}\BibitemShut {NoStop}%
\bibitem [{\citenamefont {Kremer}\ and\ \citenamefont
  {Grest}(1990)}]{kremer1990dynamics}%
  \BibitemOpen
  \bibfield  {author} {\bibinfo {author} {\bibfnamefont {K.}~\bibnamefont
  {Kremer}}\ and\ \bibinfo {author} {\bibfnamefont {G.~S.}\ \bibnamefont
  {Grest}},\ }\bibfield  {title} {\bibinfo {title} {Dynamics of entangled
  linear polymer melts: A molecular-dynamics simulation},\ }\href@noop {}
  {\bibfield  {journal} {\bibinfo  {journal} {The Journal of Chemical Physics}\
  }\textbf {\bibinfo {volume} {92}},\ \bibinfo {pages} {5057} (\bibinfo {year}
  {1990})}\BibitemShut {NoStop}%
\bibitem [{\citenamefont {Rotne}\ and\ \citenamefont
  {Prager}(1969)}]{rotne1969variational}%
  \BibitemOpen
  \bibfield  {author} {\bibinfo {author} {\bibfnamefont {J.}~\bibnamefont
  {Rotne}}\ and\ \bibinfo {author} {\bibfnamefont {S.}~\bibnamefont {Prager}},\
  }\bibfield  {title} {\bibinfo {title} {Variational treatment of hydrodynamic
  interaction in polymers},\ }\href@noop {} {\bibfield  {journal} {\bibinfo
  {journal} {The Journal of Chemical Physics}\ }\textbf {\bibinfo {volume}
  {50}},\ \bibinfo {pages} {4831} (\bibinfo {year} {1969})}\BibitemShut
  {NoStop}%
\bibitem [{\citenamefont {Yamakawa}(1970)}]{yamakawa1970transport}%
  \BibitemOpen
  \bibfield  {author} {\bibinfo {author} {\bibfnamefont {H.}~\bibnamefont
  {Yamakawa}},\ }\bibfield  {title} {\bibinfo {title} {Transport properties of
  polymer chains in dilute solution: hydrodynamic interaction},\ }\href@noop {}
  {\bibfield  {journal} {\bibinfo  {journal} {The Journal of Chemical Physics}\
  }\textbf {\bibinfo {volume} {53}},\ \bibinfo {pages} {436} (\bibinfo {year}
  {1970})}\BibitemShut {NoStop}%
\bibitem [{\citenamefont {Geyer}\ and\ \citenamefont
  {Winter}(2009)}]{geyer2009n}%
  \BibitemOpen
  \bibfield  {author} {\bibinfo {author} {\bibfnamefont {T.}~\bibnamefont
  {Geyer}}\ and\ \bibinfo {author} {\bibfnamefont {U.}~\bibnamefont {Winter}},\
  }\bibfield  {title} {\bibinfo {title} {An o (n 2) approximation for
  hydrodynamic interactions in brownian dynamics simulations},\ }\href@noop {}
  {\bibfield  {journal} {\bibinfo  {journal} {The Journal of Chemical Physics}\
  }\textbf {\bibinfo {volume} {130}},\ \bibinfo {pages} {114905} (\bibinfo
  {year} {2009})}\BibitemShut {NoStop}%
\bibitem [{\citenamefont {Miao}\ \emph {et~al.}(2017)\citenamefont {Miao},
  \citenamefont {Young},\ and\ \citenamefont {Sing}}]{miao2017iterative}%
  \BibitemOpen
  \bibfield  {author} {\bibinfo {author} {\bibfnamefont {L.}~\bibnamefont
  {Miao}}, \bibinfo {author} {\bibfnamefont {C.~D.}\ \bibnamefont {Young}},\
  and\ \bibinfo {author} {\bibfnamefont {C.~E.}\ \bibnamefont {Sing}},\
  }\bibfield  {title} {\bibinfo {title} {An iterative method for hydrodynamic
  interactions in brownian dynamics simulations of polymer dynamics},\
  }\href@noop {} {\bibfield  {journal} {\bibinfo  {journal} {The Journal of
  Chemical Physics}\ }\textbf {\bibinfo {volume} {147}},\ \bibinfo {pages}
  {024904} (\bibinfo {year} {2017})}\BibitemShut {NoStop}%
\bibitem [{\citenamefont {Young}\ \emph {et~al.}(2018)\citenamefont {Young},
  \citenamefont {Marvin},\ and\ \citenamefont
  {Sing}}]{young2018conformationally}%
  \BibitemOpen
  \bibfield  {author} {\bibinfo {author} {\bibfnamefont {C.~D.}\ \bibnamefont
  {Young}}, \bibinfo {author} {\bibfnamefont {M.}~\bibnamefont {Marvin}},\ and\
  \bibinfo {author} {\bibfnamefont {C.~E.}\ \bibnamefont {Sing}},\ }\bibfield
  {title} {\bibinfo {title} {Conformationally averaged iterative brownian
  dynamics simulations of semidilute polymer solutions},\ }\href@noop {}
  {\bibfield  {journal} {\bibinfo  {journal} {The Journal of chemical physics}\
  }\textbf {\bibinfo {volume} {149}},\ \bibinfo {pages} {174904} (\bibinfo
  {year} {2018})}\BibitemShut {NoStop}%
\bibitem [{\citenamefont {Young}\ and\ \citenamefont
  {Sing}(2019{\natexlab{b}})}]{young2019simulation}%
  \BibitemOpen
  \bibfield  {author} {\bibinfo {author} {\bibfnamefont {C.~D.}\ \bibnamefont
  {Young}}\ and\ \bibinfo {author} {\bibfnamefont {C.~E.}\ \bibnamefont
  {Sing}},\ }\bibfield  {title} {\bibinfo {title} {Simulation of semidilute
  polymer solutions in planar extensional flow via conformationally averaged
  brownian noise},\ }\href@noop {} {\bibfield  {journal} {\bibinfo  {journal}
  {The Journal of Chemical Physics}\ }\textbf {\bibinfo {volume} {151}},\
  \bibinfo {pages} {124907} (\bibinfo {year} {2019}{\natexlab{b}})}\BibitemShut
  {NoStop}%
\bibitem [{\citenamefont {Katsarou}\ \emph {et~al.}(2020)\citenamefont
  {Katsarou}, \citenamefont {Tsalikis}, \citenamefont {Tsalikis},\ and\
  \citenamefont {Mavrantzas}}]{Katsarou2020}%
  \BibitemOpen
  \bibfield  {author} {\bibinfo {author} {\bibfnamefont {A.~F.}\ \bibnamefont
  {Katsarou}}, \bibinfo {author} {\bibfnamefont {A.~J.}\ \bibnamefont
  {Tsalikis}}, \bibinfo {author} {\bibfnamefont {D.~G.}\ \bibnamefont
  {Tsalikis}},\ and\ \bibinfo {author} {\bibfnamefont {V.~G.}\ \bibnamefont
  {Mavrantzas}},\ }\bibfield  {title} {\bibinfo {title} {{Dynamic Heterogeneity
  in Polymer Blends}},\ }\href {https://doi.org/10.1007/978-3-642-36199-9_69-1}
  {\bibfield  {journal} {\bibinfo  {journal} {Polymers}\ }\textbf {\bibinfo
  {volume} {12}},\ \bibinfo {pages} {752} (\bibinfo {year} {2020})}\BibitemShut
  {NoStop}%
\bibitem [{\citenamefont {Shenoy}\ \emph {et~al.}(2016)\citenamefont {Shenoy},
  \citenamefont {Rao},\ and\ \citenamefont {Schroeder}}]{Shenoy2016}%
  \BibitemOpen
  \bibfield  {author} {\bibinfo {author} {\bibfnamefont {A.}~\bibnamefont
  {Shenoy}}, \bibinfo {author} {\bibfnamefont {C.~V.}\ \bibnamefont {Rao}},\
  and\ \bibinfo {author} {\bibfnamefont {C.~M.}\ \bibnamefont {Schroeder}},\
  }\bibfield  {title} {\bibinfo {title} {{Stokes trap for multiplexed particle
  manipulation and assembly using fluidics}},\ }\href@noop {} {\bibfield
  {journal} {\bibinfo  {journal} {Proceedings of the National Academy of
  Sciences of the United States of America}\ }\textbf {\bibinfo {volume}
  {113}},\ \bibinfo {pages} {3976} (\bibinfo {year} {2016})}\BibitemShut
  {NoStop}%
\bibitem [{\citenamefont {{Hern{\'{a}}ndez Cifre}}\ and\ \citenamefont
  {{Garc{\'{i}}a De La Torre}}(2001)}]{HernandezCifre2001}%
  \BibitemOpen
  \bibfield  {author} {\bibinfo {author} {\bibfnamefont {J.~G.}\ \bibnamefont
  {{Hern{\'{a}}ndez Cifre}}}\ and\ \bibinfo {author} {\bibfnamefont
  {J.}~\bibnamefont {{Garc{\'{i}}a De La Torre}}},\ }\bibfield  {title}
  {\bibinfo {title} {{Kinetic aspects of the coil-stretch transition of polymer
  chains in dilute solution under extensional flow}},\ }\href
  {https://doi.org/10.1063/1.1410379} {\bibfield  {journal} {\bibinfo
  {journal} {Journal of Chemical Physics}\ }\textbf {\bibinfo {volume} {115}},\
  \bibinfo {pages} {9578} (\bibinfo {year} {2001})}\BibitemShut {NoStop}%
\bibitem [{\citenamefont {Peddireddy}\ \emph
  {et~al.}(2020{\natexlab{b}})\citenamefont {Peddireddy}, \citenamefont {Lee},
  \citenamefont {Schroeder},\ and\ \citenamefont
  {Robertson-Anderson}}]{Peddireddy2020a}%
  \BibitemOpen
  \bibfield  {author} {\bibinfo {author} {\bibfnamefont {K.~R.}\ \bibnamefont
  {Peddireddy}}, \bibinfo {author} {\bibfnamefont {M.}~\bibnamefont {Lee}},
  \bibinfo {author} {\bibfnamefont {C.~M.}\ \bibnamefont {Schroeder}},\ and\
  \bibinfo {author} {\bibfnamefont {R.~M.}\ \bibnamefont
  {Robertson-Anderson}},\ }\bibfield  {title} {\bibinfo {title} {{Viscoelastic
  properties of ring-linear DNA blends exhibit non-monotonic dependence on
  blend composition}},\ }\href
  {https://doi.org/10.1103/PhysRevResearch.2.023213} {\bibfield  {journal}
  {\bibinfo  {journal} {Physical Review Research}\ }\textbf {\bibinfo {volume}
  {2}},\ \bibinfo {pages} {023213} (\bibinfo {year}
  {2020}{\natexlab{b}})}\BibitemShut {NoStop}%
\end{thebibliography}
%

\end{document}